\documentclass[aps,pre,reprint, amsmath, amssymb]{revtex4-1}

\usepackage{bm}
\usepackage[retainorgcmds]{IEEEtrantools}
\usepackage{graphicx}
\usepackage{mathrsfs}
\usepackage{amsmath}
\usepackage{amssymb}
\usepackage{color}
\usepackage{amsfonts}

\newcommand{\ud}{\,\mathrm{d}}

\DeclareMathOperator{\tr}{tr}

\newcommand{\trans}{^{\text{T}}}

\begin{document}

\title{Fluctuations of the heat exchanged between two quantum spin chains}
\date{\today}
\author{Gabriel T. Landi}
\email{gtlandi@gmail.com}
\affiliation{Universidade Federal do ABC,  09210-580 Santo Andr\'e, Brazil}
\author{Dragi Karevski}
\affiliation{Institut Jean Lamour, Department P2M, Groupe de Physique Statistique, Universit\'e de Lorraine, CNRS, B.P. 70239, F-54506 Vandoeuvre les Nancy Cedex, France}

\begin{abstract}

The statistics of the heat exchanged between two quantum XX spin chains prepared at different temperatures is studied within the assumption of weak coupling. 
This provides simple formulas for  the average heat and its corresponding characteristic function, from which the  probability distribution may be computed numerically. 
These formulas  are valid for arbitrary sizes and therefore allow us to analyze the role of the thermodynamic limit in this non-equilibrium setting. 
It is found that all thermodynamic quantities are extremely sensitive to the quantum phase transition of the XX chain.


\end{abstract}
\maketitle{}

%
%
%
%

\section{Introduction}

When studying the thermodynamic properties of small out of equilibrium systems  (quantum or classical), it is found that quantities such as heat flux, work and entropy production may widely fluctuate in the course of time or between each realization of an experiment. 
Over the past two decades, researchers have discovered that the probability distributions governing these fluctuations obey certain symmetry relations 
which touch into the nature of irreversibility  and the arrow of time \cite{DelCampo2014,Batalhao2015}. 
These relations are now generally known as fluctuation theorems \cite{
Evans1993,*Evans1994,*Evans2002,*Williams2008,*Evans2011b,
Gallavotti1995,*Gallavotti1995b,
Sivak2012a,
Crooks1998,*Crooks2000,*Crooks2008,
Jarzynski1997a,*Jarzynski1997,*Jarzynski1999a,*Jarzynski2000,*Jarzynski2001a,*Jarzynski2011a,
Kurchan1998,
Maes1999,
Lebowitz1999,
Jarzynski2004a,
Hoppenau2013,
Sinitsyn2011,
Sivak2012a,
Tasaki2000,
Kurchan2000,
Mukamel2003,
Monnai2005,
Teifel2007,*Teifel2011,
Talkner2007,*Talkner2009,*Talkner2013a,
Gelin2008,
Vaikuntanathan2009,
Esposito2009,
Campisi2011,
Cohen2012a,
Olshanii2012a,
Chetrite2012,
Goold2014b}.
Historically,  they were initially developed for classical systems obeying either Hamiltonian or stochastic dynamics \cite{
Evans1993,*Evans1994,*Evans2002,*Williams2008,*Evans2011b,
Gallavotti1995,*Gallavotti1995b,
Sivak2012a,
Crooks1998,*Crooks2000,*Crooks2008,
Jarzynski1997a,*Jarzynski1997,*Jarzynski1999a,*Jarzynski2000,*Jarzynski2001a,*Jarzynski2011a,
Kurchan1998,
Maes1999,
Lebowitz1999,
Jarzynski2004a,
Hoppenau2013,
Sinitsyn2011,
Sivak2012a}.
Their validity was also extensively verified experimentally \cite{VanZon2004a,Douarche2005,Gomez-Solano2011,Ciliberto2013}.
Conversely, their development for quantum systems is much more recent \cite{
Tasaki2000,
Kurchan2000,
Mukamel2003,
Monnai2005,
Teifel2007,*Teifel2011,
Talkner2007,*Talkner2009,*Talkner2013a,
Gelin2008,
Vaikuntanathan2009,
Esposito2009,
Campisi2011,
Cohen2012a,
Olshanii2012a,
Chetrite2012,
Goold2014b}, particularly their experimental verification \cite{Batalhao2014}.

The probability distributions  or characteristic functions of thermodynamic quantities are the central objects in this theory. 
Hence, in order to gain additional insight into this  problem, it is important to compute them in specific models, which are tractable either theoretically \cite{Bena2005,Cleuren2006,*Cleuren2007,Crooks2007,Chatelain,*Chatelain2007,Dorosz2008a,Piscitelli2009,Dorner2012,Cornu2013,Brunelli2014,Fusco2014a} or experimentally \cite{VanZon2004a,Douarche2005,Gomez-Solano2011,Ciliberto2013,Batalhao2014,Batalhao2015}.
In particular, Heat fluctuations in quantum systems has recently been the subject of a large number of studies \cite{Saito2007,Nicolin2011,Kundu2011,Panasyuk2012,Fogedby2014,Piscitelli2008,Sughiyama2008,Corberi2013,Lahiri,Gasparinetti2014,Agarwalla2014,Kim2014a,Goold2014a,Collura2014,Akagawa2009,Gomez-Marin2006}
and the purpose of our work is to discuss this problem on an exactly solvable case, namely the quantum XX spin chain.  

The setup considered here is the common thermalization of two bodies, which were separately prepared in equilibrium at  different temperatures $T_1$ and $T_2$ and then put in contact through a diabatic wall (no work is performed during the entire process) \cite{Callen1985}.  
According to the second law of thermodynamics, in the formulation of Clausius, when the two bodies are macroscopic, heat should flow from the hotter to the colder body until thermal equilibrium is reached at a common temperature $T^*$. 
If the two bodies are equal, the total heat exchanged between them  during this process should be given by
\begin{equation}\label{thermo_heat}
\mathcal{Q} = \frac{U(T_2) - U(T_1)}{2}
\end{equation}
where $U(T)$ is the internal energy of the body. This is the basic prediction of thermodynamics. 

However, the situation is quite different if the two systems are not macroscopically large.  Due to wide fluctuations,
the observed values of the heat exchanged  during the process
 (empirically defined as the energy difference between the initial and the final state) may fluctuate substantially with respect to the thermodynamical prediction.
In extreme cases it is in principle possible to observe a heat flow in the opposite direction, from the colder system to the hotter.
Let us denote by $P(Q)$ the probability that a certain amount of heat $Q$ entered system 1 after a time $t$ (a proposal to experimentally measure $P(Q)$ was recently given in Ref.~\cite{Goold2014a}).
 In Ref.~\cite{Jarzynski2004a} Jarzynski and W\'ojcik showed that $P(Q)$ obeys the following fluctuation theorem: 
 \begin{equation}\label{FT}
 \frac{P(Q)}{P(-Q)} = e^{\Delta \beta \;Q}
 \end{equation}
where $\Delta \beta = \frac{1}{T_1} - \frac{1}{T_2}$. 
This relation was derived under the assumption of \emph{weak coupling}; i.e., when the interaction energy between the two systems is negligibly small compared to the "bulk" individual energies. It implies in particular that the energy which enters system 1 is, to a good approximation, the same as the energy which left system 2. 

Eq.~(\ref{FT}) may be physically interpreted as saying that it is exponentially more likely that heat will flow in the right direction.
Of course, if we assume that energy is an extensive quantity, for macroscopic bodies the probability of heat flowing in the opposite direction becomes negligible. 
From Eq.~(\ref{FT}) it also follows that $Q$ obeys a non-equilibrium equality:
\begin{equation}\label{equality}
\langle e^{-\Delta \beta Q} \rangle_t = 1
\end{equation}
 Using Jensen's inequality and assuming that $T_2 > T_1$, this leads to
 \begin{equation}\label{Qpositive}
 \langle Q \rangle_t \geq 0\; ,
 \end{equation}
such that the second law is recovered as an average statement valid also for microscopic bodies. 
 Another interesting fact about Eq.~(\ref{FT}) is that it holds whether or not the two systems actually thermalize. 
In fact, since the two bodies  are isolated from the rest of the universe, the time evolution is unitary and  
depending on the system in question, it is possible that the heat will oscillate indefinitely, back and forth between the two bodies. 

The distribution of heat is currently of importance in quantum thermodynamics due to its intimate relation to Landauer's principle and its consequences to quantum information \cite{Landauer1975,Goold2014b}. 
But, as with other distributions in quantum thermodynamics, it is not readily computable for many-body problems. 
In this paper we will study the two bodies thermalization setup for the integrable case of two finite quantum XX  spin chains connected by an XX coupling. 
This model is exactly diagonalizable in terms of Fermionic operators. 
Notwithstanding, computing the distribution $P(Q)$ for arbitrary sizes remains a difficult task due to the enormous number of transitions inherent to the many-body problem (in a sense, it amounts to a bookkeeping problem).
Instead we shall adopt a different route and exploit the fact that, if the size of the two chains is moderately large,  the coupling between them will be comparatively much smaller such that we may treat the interaction potential perturbatively (this approximation is scrutinized in detail in Appendix~\ref{sec:alt}).

It is shown that with this approximation the characteristic function of $Q$ becomes a simple product of terms,  from which 
the probability $P(Q)$ may  be computed numerically very easily for any desired sizes. 
Moreover, the average heat $\langle Q \rangle_t$ is found to be described by a  simple formula with a clear physical interpretation. 
It will be shown that all thermodynamic quantities are extremely sensitive to the quantum phase transition that the XX chain undergoes as a function of the magnetic field \cite{Niemeijer1967}. 
For completeness, in Appendix~\ref{sec:work} we also study  the distribution of the work that must be performed by an external agent in turning on the interaction at time $t=0$, a problem which was recently investigated for small system sizes in Ref.~\cite{Apollaro2014}.

%
%
%
%

\section{\label{sec:formal}Formal Framework}

We begin with two systems with Hamiltonians $H_1$ and $H_2$ which, for simplicity, have the same energy spectrum:
\begin{IEEEeqnarray}{rCl}
\label{eig1} H_1 |n_1\rangle &=& E_{n_1} |n_1\rangle \\[0.2cm]
\label{eig2} H_2 |n_2\rangle &=& E_{n_2} |n_2\rangle
\end{IEEEeqnarray}
The two systems have been prepared in thermal equilibrium at different temperatures, $T_1$ and $T_2$, so that 
the  initial density matrix of the composite system is given as a product  state
\begin{equation}\label{rho_thermal}
\rho_\text{th} = \frac{e^{-\beta_1 H_1}}{Z_1}\otimes \frac{e^{-\beta_2 H_2}}{Z_2}
\end{equation}
where $\beta_{1,2} = 1/T_{1,2}$ (the Bolztmann constant is set to $k_B = 1$) and $Z_{1,2}$ are the corresponding partition functions. 


At $t=0^-$ we then unplug the systems from their reservoirs and measure the observables $H_1$ and $H_2$, obtaining the state $|\bm{n} \rangle = |n_1,n_2\rangle$   with probability 
\begin{equation}\label{pn}
p_{\bm{n}} = \langle \bm{n}| \rho_\text{th} | \bm{n}\rangle = \frac{e^{-\beta_1 E_{n_1}}}{Z_1} \frac{e^{-\beta_2 E_{n_2}}}{Z_2}
\end{equation}
Immediately after this measurement, at $t = 0^+$, we turn on an interaction $V$ between the two systems (i.e., we perform a quantum quench) such that the total Hamiltonian passes from 
\begin{equation}\label{H0}
H_0 = H_1 + H_2 
\end{equation}
at $t = 0^-$ to 
\begin{equation}\label{H}
H = H_1 + H_2 + V
\end{equation}
at $t = 0^+$
\footnote{This process modifies a parameter in the Hamiltonian and therefore requires  the expenditure of work.
Since it occurs instantaneously and since the composite system is isolated, the average work performed must be 
\unexpanded{$
\langle W \rangle = \langle V \rangle_0 = \tr(V \rho_\text{th})
$}
}.
Next we allow the composite system to evolve unitarily with Hamiltonian $H$ up to a certain time $t$, when we measure the observables $H_1$ and $H_2$ one more time. 
Suppose that the state obtained from this measurement is $|\bm{m} \rangle = |m_1,m_2\rangle$. 
Then the total change in the energy of each subsystem is, by definition, 
\begin{IEEEeqnarray}{rCl}
\label{Q1def} Q_1 &=& E_{m_1} - E_{n_1}  \\[0.2cm]
\label{Q2def} Q_2 &=& E_{m_2} - E_{n_2} 
\end{IEEEeqnarray}


The quantities $Q_{1,2}$ are random variables described by probability distributions $P_1(\mathcal{Q})$ and $P_2(\mathcal{Q})$. 
A formal expression for these distributions may be obtained by noting that $|\langle   \bm{m}|e^{-i H t}|\bm{n} \rangle|^2$ is the conditional probability that the system is found  in the state $|\bm{m}\rangle$ at time $t$, given that it was initially at $|\bm{n}\rangle$ at time $0$. 
We therefore have
\begin{equation}\label{PQ_def}
P_{1}(\mathcal{Q}) = \sum\limits_{\bm{n},\bm{m}} |\langle \bm{m}|e^{-i H t}|\bm{n}\rangle|^2 p_{\bm{n}} \; \delta[q - (E_{m_{1}} - E_{n_{1}})].
\end{equation}
and a similar definition for $P_{2}(\mathcal{Q})$.
It is much more convenient, however, to work with the characteristic function $F_1(r) = \langle e^{i r Q_1} \rangle_t$.
From Eq.~(\ref{PQ_def}) it can be shown that
\begin{IEEEeqnarray}{rCl}
F_1(r) \equiv \langle e^{i r Q_1} \rangle_t &=& \tr\left\{e^{i H t} e^{i r H_1}e^{-i H t} e^{-i r H_1} \rho_\text{th} \right\}\nonumber \\[0.2cm]
\label{F}  &=&\tr\left\{ e^{i r H_1(t)} e^{-i r H_1} \rho_\text{th} \right\},\;  
\end{IEEEeqnarray}
where we defined  the Heisenberg representation of an operator $A$ as $A(t) =  e^{i H t} A e^{-i H t}$. 
We therefore see that the characteristic function is a time-ordered correlation function evaluated at the initial (thermal) state \cite{Talkner2007}.
Under weak coupling, when the fluctuation theorem~(\ref{FT}) is valid, it follows by definition that $F_1(i\Delta \beta) = 1$.

The average heat may be obtained by the definition 
\begin{equation}\label{av_def}
\langle Q_{1,2} \rangle_t = \int d\mathcal{Q} \; \mathcal{Q} P_{1,2}(\mathcal{Q})
\end{equation}
or by expanding $F_1(r)$ in a power series in $r$. In either case one obtains
\begin{equation}\label{QaveDef}
\langle Q_1 \rangle = \tr[(H_1(t) - H_1) \rho_\text{th}]
\end{equation}
again with a similar formula for $\langle Q_2 \rangle$. 
From this formula it appears to be  possible to associate the random variable $Q_1$ with the quantum mechanical operator  $\delta H_1(t) = H_1(t) - H_1$. 
However, such an association is incorrect since to access $Q_1$ requires two energy measurements  \cite{Allahverdyan2005,Dorosz2008a,Talkner2007}  \footnote{Notice that a recent proposal suggested that work (and consequently heat) can be measured by performing a generalized quantum measurement at a single time \cite{Chiara2015,Roncaglia2014a}}. This reflects the fact that $Q_1$ is not a property of the system, but rather the outcome of a process. 
This fact also becomes evident when computing higher moments of $Q_1$.
Due to the fact that $[H_1(t),H_1] \neq 0$, it follows that higher moments of $Q_1$ cannot be associated with higher moments of $\delta H_1(t)$.

Notice that since the evolution is unitary, with total Hamiltonian $H(t)=H=H_1+H_2 + V$, we have 
\begin{equation}\label{conserve}
\langle Q_1 \rangle_t + \langle Q_2 \rangle_t = \langle V \rangle_0 - \langle V \rangle_t\; .
\end{equation}
The term $-\langle V \rangle_t$ is the average work that must be performed in turning off the interaction at time $t$ \cite{Akagawa2009}. 
Thus, the right-hand side of this expression is the total work performed by the external agent in turning the interaction on and off. 
We therefore see that in general it is not correct to interpret $Q_1$ and $Q_2$ as the heat which entered each system, since part of the change may be due to the work performed by the external agent.
This will only be true in the \emph{weak coupling limit}, where $V \ll H_{1,2}$. 
In this case the right-hand side of Eq.~(\ref{conserve}) is negligible and we may therefore define $\langle Q \rangle_t := \langle Q_1 \rangle_t \simeq - \langle Q_2 \rangle_t$.
Hence, in the weak coupling limit it is correct to attribute the changes in the energy of system 1 as being due to the heat which flowed from system 2. 

It is also worth  emphasizing that the definition of heat is not tied to the fact that the initial state is thermal. 
Any energy entering system 1 from system 2, in the assumption of weak coupling, is correctly defined as heat. 
The only difference is that, if the initial states are not thermal, certain thermodynamic properties and the fluctuation theorem~(\ref{FT}) will not necessarily hold. 



%
%
%
%

\section{\label{sec:model}The model}

We now apply these concepts to a model consisting of  two identical XX quantum spin chains of size $L$  coupled together through an XX coupling. 
The total Hamiltonian is $H=H_1+H_2 +V$, where the  Hamiltonians of the individual chains $H_1$ and $H_2$ and the  interaction potential $V$ are given in terms of the usual Pauli algebra by
\begin{IEEEeqnarray}{rCl}
\label{H1}H_1 &=& \frac{h}{2} \sum\limits_{i=1}^L \sigma_i^z - \frac{J}{2} \sum\limits_{i=1}^{L-1} (\sigma_i^x \sigma_{i+1}^x + \sigma_i^y \sigma_{i+1}^y) \\[0.2cm]
\label{H2}H_2 &=& \frac{h}{2} \sum\limits_{i=L+1}^{2L} \sigma_i^z - \frac{J}{2} \sum\limits_{i=L+1}^{2L-1} (\sigma_i^x \sigma_{i+1}^x + \sigma_i^y \sigma_{i+1}^y) \\[0.2cm]
\label{V}V &=& \frac{g_0}{2} (\sigma_L^x \sigma_{L+1}^x + \sigma_L^y \sigma_{L+1}^y)
\end{IEEEeqnarray}

%
%
The standard Fermionization of $H$ is accomplished through the Jordan-Wigner mapping
\begin{equation}\label{cops}
c_n = \left[ \prod\limits_{j=1}^{n-1} (-\sigma_j^z) \right] \sigma_n^-
\end{equation}
where $\sigma_n^\pm = (\sigma_n^x \pm i \sigma_n^y)/2$ and $n = 1, \ldots, 2L$. 
The Fermi operators $c_n$  satisfy the  canonical anticommutation relations
\begin{equation}\label{com}
\left\{ c_n, c_m^\dagger \right\} = \delta_{n,m}\; .
\end{equation}
In terms of these operators Eqs.~(\ref{H1})-(\ref{V}) become
\begin{IEEEeqnarray}{rCl}
\label{H1c} H_1 &=& h \sum\limits_{n=1}^L c_n^\dagger c_n - J \sum\limits_{n=1}^{L-1} (c_n^\dagger c_{n+1} + c_{n+1}^\dagger c_n)  \\[0.2cm]
\label{H2c} H_2 &=& h \sum\limits_{n=L+1}^{2L} c_n^\dagger c_n - J \sum\limits_{n=L+1}^{2L-1} (c_n^\dagger c_{n+1} + c_{n+1}^\dagger c_n) \\[0.2cm]
\label{Vc} V &=& g_0 (c_L^\dagger c_{L+1} + c_{L+1}^\dagger c_L) 
\end{IEEEeqnarray}
where in $H_1$ and $H_2$ we omitted an irrelevant constant term $Lh/2$. 

The  Hamiltonians $H_1$ and $H_2$ may be individually diagonalized by a Bogoliubov transformation, introducing  the new set of Fermionic operators 
\begin{equation}\label{b-def}
a_k = \sum\limits_{n=1}^L x_{n,k} c_n\; ,
\qquad b_{k} = \sum\limits_{n=1}^L x_{n,k} c_{n+L}
\end{equation}
where
\begin{equation}\label{xkl}
x_{n,k} = \sqrt{\frac{2}{L+1}} \sin(nk)
\end{equation}
with $n = 1,\ldots,L$ and $k(L+1) = \pi,2\pi,\ldots,L\pi$.
We then obtain
\begin{equation}\label{H12}
H_1 = \sum\limits_{k} \lambda_k a_k^\dagger a_k\; ,
\qquad H_2 = \sum\limits_{k} \lambda_k b_{k}^\dagger b_{k}
\end{equation}
where the single particle energy-spectrum is given by
\begin{equation}\label{lambda}
\lambda_k = h - 2 J \cos k \; .
\end{equation}
This dispersion relation shows a gap for $h>h_c\equiv 2J$ indicating the appearance of a quantum phase transition between a gapless super-fluid phase for $h< 2J$ and a gapped Mott-insulating phase for $h>2J$ \cite{Niemeijer1967,Karevski2000}. 

In terms of the $a$ and $b$ Fermi operators the interaction energy $V$ in Eq.~(\ref{Vc}) can be recast as
\begin{equation}\label{Vb}
V = \sum\limits_{k,q} G_{k,q} (a_k^\dagger b_q + b_q^\dagger a_k)
\end{equation}
with 
\begin{equation}\label{Gkl}
G_{k,q} = g \sin(Lk)\sin(q) 
\end{equation}
where 
\begin{equation}\label{g}
g = \frac{2g_0}{L+1}  \; .
\end{equation}

At $t>0$, the total Hamiltonian is therefore given by
\begin{equation}\label{Htot}
H = \sum\limits_k \lambda_k(a_k^\dagger a_k + b_k^\dagger b_k) + \sum\limits_{k,q} G_{k,q} (a_k^\dagger b_q + b_q^\dagger a_k)\; .
\end{equation}
The problem has  thus been reduced to a quadratic Hamiltonian with two different types of fermions (representing the two chains). 
The last term in this formula describes scattering  processes occurring at the junction sites. 
Each process represents the annihilation of a b-fermion in chain 2 with momentum $q$ and the creation of an a-fermion in chain 1 with momentum $k$, and vice-versa, each event  occurring  with amplitude $G_{k,q}$. 
This event decreases the energy of chain 2  by $\lambda_q$ while increasing the energy of chain one  by $\lambda_k$. 
Hence, the total energy of the two chains,  $H_0 = H_1+H_2$, changes by $\lambda_k - \lambda_q$ and is therefore not a conserved quantity. 
This can also be seen directly from the fact that 
\begin{equation}
[H_1+H_2,H]=\sum_{k\neq q} (\lambda_k-\lambda_q) G_{k,q} [a_k^+b_q-b_q^+a_k]
\end{equation}
Alternatively, we may say that in each scattering process some energy is retained in, or extracted from, the interaction potential $V$. 
The only transitions in which the energy of the individual chains is conserved are those where $k = q$.

However, notice also that $G_{k,q}$ scales as $g_0/L$ which will therefore be small if either $g_0$ is small or $L$ is large. 
Hence, as the size of the chains increase, the energy retained in $V$ during these transitions diminishes when compared with the typical energies of the system. 
This is precisely the idea of \emph{weak coupling}
 and is  in agreement with our perception that $V$ is a boundary interaction and is thus much smaller than the bulk Hamiltonians $H_{1,2}$. 

The next step is to find the time propagator $U(t) = e^{-i H t}$. 
This may be done exactly since $H$ is quadratic in Fermi operators. However, the exact resulting formulas are extremely cumbersome to work with.
Instead, motivated by the fact that the interaction term is small, we will derive an approximate formula for $U(t)$, that will hold when either $g_0$ is small and/or $L$ is large. 

Moving to the interaction picture, the potential $V_I(t) = e^{i H_0 t} V e^{-i H_0 t}$ becomes
\begin{equation}\label{VI}
V_I(t) = \sum\limits_{k,q} G_{k,q} \Big\{a_k^\dagger b_q e^{i(\lambda_k - \lambda_q)t} + b_q^\dagger a_k e^{-i(\lambda_k-\lambda_q)t}\Big\}
\end{equation}
It can be seen that the terms with $k\neq q$ are rapidly oscillatory, and average to zero at long times. 
Hence, they may be neglected provided $G_{k,q}$ is small and provided that the time scales of interest are larger than the typical separation $\lambda_k -\lambda_q$ of the energy levels. 
Both requirements are true in this case. 
The former because $G_{k,q}\propto 1/L$ and the latter because, as will be seen below, the typical time scales of interest for the thermodynamic quantities are of order $L$, whereas the smallest energy difference $\lambda_k - \lambda_q$ is of order $1/L$.
We therefore conclude that when $g$ is small, the Hamiltonian in Eq.~(\ref{Htot}) may well be replaced in the long time limit by the approximate expression  
\begin{equation}\label{Hk}
H = \sum\limits_k H_k =  \sum\limits_k \left\{ \lambda_k(a_k^\dagger a_k + b_k^\dagger b_k) + G_{k} (a_k^\dagger b_k + b_k^\dagger a_k)\right\}
\end{equation}
taking into account only the diagonal transition rates $G_k = G_{k,k}$. 
In Appendix~\ref{sec:alt} the accurateness of this approximation is tested by comparing it with the numerically exact results obtained by the exact diagonalization of  Eq.~(\ref{Htot}). 
It is shown that the agreement is extremely good, even for moderately large $L$, and improves with increasing $L$ and/or decreasing $g_0$.

%
%
%
%


The initial thermal state is a tensor product state that may be factored as a product of $L$ terms, each associated to the mode $k$:
\[
\rho_\text{th} = \prod\limits_k \rho_{\text{th},k}, \qquad \rho_{\text{th},k} = \frac{e^{-\beta_1 \lambda_k a_k^\dagger a_k - \beta_2 \lambda_k b_k^\dagger b_k}}{Z_k}
\]
with  normalization factor 
\[
Z_k = (1+e^{-\beta_1 \lambda_k}) (1+e^{-\beta_2 \lambda_k})\; .
\]
Similarly the Hamiltonian in Eq.~(\ref{Hk}) factors into a sum of $L$ commuting terms. 
Thus,  the entire dynamics factors into $L$ independent subspaces where the $a_k$ of chain 1 are only coupled  to the mode $b_k$ of chain 2. 
Each subspace has dimension 4 and is spanned by the vectors $|0,0\rangle$, $|k,0\rangle = a_k^\dagger|0,0\rangle$, $|0,k\rangle = b_k^\dagger |0,0\rangle$ and $|k,k\rangle = a_k^\dagger b_k^\dagger |0,0\rangle$ (this is similar to the situation treated in Ref.~\cite{Lieb1964}).
In this basis, apart from a phase factor $e^{-i \lambda_k t}$ that we may omit, the dynamics is generated by
\begin{equation}\label{time_propagator}
e^{-i H_k t} =  \begin{bmatrix}
e^{i \lambda_k t} & 0 & 0 & 0 \\[0.2cm]
0 & \cos(G_k t) & -i \sin(G_k t) & 0 \\[0.2cm]
0 & -i \sin(G_k t) & \cos(G_k t) & 0 \\[0.2cm]
0 & 0 &0  & e^{-i \lambda_k t} 
\end{bmatrix}
\end{equation}

As a consequence, the reduced density matrix associated to the mode $k$ at time $t$, $\rho_k(t) = e^{-i H_k t} \rho_{\text{th},k} e^{i H_k t}$, has a block diagonal structure which is non trivial only in the sector $\{|k,0\rangle, |0,k\rangle\}$: 
\begin{equation}\label{rho_kt}
\rho_k(t) =\begin{pmatrix}
h_k^1h_k^2 &0 & 0 & 0 \\[0.2cm]
0 &n_k^1h_k^2 + f_k(t)&g_k^*(t) & 0 \\[0.2cm]
0 & g_k(t)&  n_k^2h_k^1 - f_k(t)& 0 \\[0.2cm]
0 & 0 & 0 &  n_k^1  n_k^2 
\end{pmatrix}
\end{equation}
where 
\begin{equation}\label{nk}
n_k^\alpha=\frac{1}{e^{\lambda_k/T_\alpha}+1}\; , \quad h_k^\alpha=1-n_k^\alpha  \qquad \alpha = 1,2
\end{equation}
are the initial occupation numbers of particles and holes and where 
\begin{IEEEeqnarray}{rCl}
\label{aa}  f_k(t)&=&  ( n_k^2 -  n_k^1)\sin^2(G_k t) \\[0.2cm]
\label{ab}  g_k(t) &=& i(n_k^2-n_k^1)\sin(2G_kt) \; .
\end{IEEEeqnarray} 
We see from (\ref{rho_kt}), (\ref{aa}) and (\ref{ab}) that the modes oscillate between the left and right sides with a mode-dependent period $t_k^*=\frac{\pi}{|G_k|}\simeq \frac{\pi L}{2 g_0 \sin^2 k}$, such that the fastest oscillations occur with a typical period of order $L$, while the slowest with a typical period of order $L^3$. 

The occupation numbers and correlations of the $a$ and $b$ fermions are readily computed from Eq.~(\ref{rho_kt}):
\begin{IEEEeqnarray}{rCl}
 \langle a_k^\dagger a_k \rangle_t &=& n_k^1 +f_k(t) \\[0.2cm]
 \langle b_k^\dagger b_k \rangle_t &=& n_k^2 - f_k(t) \\[0.2cm]
 \langle a_k^\dagger b_k \rangle_t &=& ig_k(t)
\end{IEEEeqnarray} 
The total number of excitations with momentum $k$ is therefore conserved, as expected.

Finally, it is worth mentioning that the long-time average of the reduced density matrix,
\[
\bar{\rho_k}(t) \equiv \lim_{T\rightarrow \infty} \frac{1}{T}\int\limits_{t}^{t+T} \rho_k(t) dt,
\]
reduces to a stationary state where the left and right states $|k,0\rangle$, $|0,k\rangle$ have equal weights 
$\frac{n_k^1 (1-n_k^2) + n_k^2(1-n_k^1)}{2}$.



%
%
%
%

\section{\label{sec:thermo}Thermodynamic properties}

\subsection{\label{sec:non}General out-of-equilibrium properties}

Now that we have the time evolution operator we may study any thermodynamic quantity of interest. 
We will focus our discussion on the heat $Q_1$ entering the first sub-system. 
Due to the weak coupling assumption, this will be equal to $-Q_2$. 
For this reason,  we will henceforth omit any suffixes $1$ in the formulas developed in Sec.~\ref{sec:formal}.

All operators appearing in the characteristic function, Eq.~(\ref{F}), factor into products of $L$ commuting operators. Hence, $F(r)$ may be written as 
\begin{equation}\label{F_prod}
F(r) = \prod\limits_k F(k,r)
\end{equation}
This means that the total heat $Q$ may be written as a sum of independent  (\emph{not} identically distributed) random variables $Q_k$:
\begin{equation}\label{Qsum}
Q = \sum\limits_k Q_k
\end{equation}
such that $F(k,r) = \langle e^{i r Q_k} \rangle$.
Each $Q_k$ physically represents the heat exchanged between each pair of normal modes $a_k$ and $b_k$ and it is defined as positive when energy enters chain 1. 

The functions $F(k,r)$ may be readily computed  using Eq.~(\ref{rho_kt}). 
They read
\begin{equation}\label{Fk}
F(k,r) = p_k^0 + e^{i\lambda_k r} p_k^+ + e^{-i \lambda_k r} p_k^-
\end{equation}
where
\begin{IEEEeqnarray}{rCl}
\label{pkp} p_k^+  &=& n_k^2(1-n_k^1) \sin^2(G_k t)\\[0.2cm]
\label{pkm} p_k^-  &=& n_k^1(1-n_k^2) \sin^2(G_k t)\\[0.2cm]
\label{pk0} p_k^0   &=& 1-p_k^+ - p_k^-
\end{IEEEeqnarray}
Taking the inverse Fourier transform we find that this corresponds to a discrete process with density
\begin{equation}\label{pdis}
\pi_k(Q)=p_k^0 \delta(Q)+ p_k^+\delta(Q-\lambda_k)+p_k^-\delta(Q+\lambda_k)\; .
\end{equation}
When chain 1 gains an excitation from chain 2, $Q_k = \lambda_k$, which occurs with probability $p_k^+$. 
The inverse process corresponds to $Q_k = -\lambda_k$ and occurs with probability $p_k^-$. 
Finally, there is also the probability that no excitations are exchanged, which is $p_k^0$.

The probabilities $p_k^\pm$ in Eqs.~(\ref{pkp}) and (\ref{pkm}) obey the fluctuation theorem (\ref{FT}):
\[
\frac{p_k^+}{p_k^-} = e^{\Delta \beta \lambda_k}
\]
where $\Delta \beta = \beta_1 - \beta_2$. 
This result  is expected since the approximations  that led us to Eq.~(\ref{Hk}) were based on the assumption of weak-coupling. 
The fluctuation theorem is also manifested in Eq.~(\ref{Fk}) by the fact that $F(r=i \Delta \beta) = 1$ [see also Eq.~(\ref{equality})].
When $T_1 = T_2$ the characteristic function becomes real and therefore $P(\mathcal{Q})$ becomes even in $Q$. 
In this case heat is equally likely to flow in either direction. 
Conversely, when $T_2 \neq T_1$, the distribution will be asymmetric.

%
%
%
%

\subsection{\label{ssec:ave} Average Heat}
\subsubsection{Analytical results}
The average heat $Q_k$ associated to mode $k$ may be readily computed from Eq.~(\ref{pdis}):
\begin{equation}\label{Qk_ave}
\langle Q_k \rangle_t  = \lambda_k (p_k^+ - p_k^-)= \lambda_k (n_k^2 - n_k^1) \sin^2(G_k t)\; .
\end{equation}
The total heat is then simply a sum over all allowed values of $k$:
\begin{equation}\label{Qfin}
\langle Q \rangle_t  =\sum_k  \lambda_k (p_k^+ - p_k^-)= \sum_k \lambda_k (n_k^2 - n_k^1) \sin^2(G_k t)\; .
\end{equation}

In the thermodynamic limit we may replace the sum by an integral and work with the heat density defined as $q_t\equiv \lim_{L\rightarrow \infty} \langle Q \rangle_t/L$. We then find
\begin{equation}\label{qt}
q_t  = \frac{1}{\pi}\int_0^\pi \ud k\; \lambda_k (n_k^2 - n_k^1) \sin^2(g t \sin^2 k)\; .
\end{equation}
We see that the relevant time scale of the thermodynamic quantities is $1/g$. 
When $gt$ is sufficiently large, the $\sin$ term will oscillate extremely fast as $k$ is varied from $0$ to $\pi$. 
We may then approximate $\sin^2(g t \sin^2 k) \sim 1/2$, which leads to the time averaged heat density 
\begin{IEEEeqnarray}{rCl}
\label{qbar} \bar{q} &\equiv& \lim\limits_{t\rightarrow \infty} \lim\limits_{\tau\rightarrow \infty} \frac{1}{\tau}\int\limits_t^{t+\tau} q_t \ud t \nonumber \\[0.2cm]
&=& \frac{1}{2\pi}\int_0^\pi \ud k\;\lambda_k (n_k^2 - n_k^1) = \frac{u(T_2)-u(T_1)}{2}\; 
\end{IEEEeqnarray}
where $u(T)$ is the equilibrium energy density at temperature $T$ of a single XX chain. 
This is nothing but the classical thermodynamical result (\ref{thermo_heat}).

These results are valid for arbitrary initial temperatures $T_1$ and $T_2$. 
Specializing further to the case close to equilibrium, with $T_1 = T$ and  $T_2 = T+\Delta T$ where $\Delta T\ll T$, we obtain 
\begin{equation}\label{Qave}
q_t  = \frac{\Delta T}{\pi}\int_0^\pi dk\; c_k(T) \sin^2(g t \sin^2 k)\; .
\end{equation}
where 
\begin{equation}\label{ck}
c_k(T) = \lambda_k \frac{\partial n_k}{\partial T} = \left(\frac{\lambda_k}{T}\right)^2 \frac{e^{\lambda_k/T}}{ \left(e^{\lambda_k/T}+1\right)^2}
\end{equation}
is the heat capacity of mode $k$ of a single XX chain at a temperature $T$. The time average heat density reduces to 
\begin{equation}\label{Qave3}
\bar{q}  = \frac{\Delta T}{2\pi}\int_0^\pi dk\; c_k(T) = \frac{\Delta T}{2} c(T)\; ,
\end{equation}
which is precisely the thermodynamic heat density, with $c(T)$ being the specific heat of the XX chain.

The above calculations therefore show that in the thermodynamic limit one recovers the expectations of classical thermodynamics in the long-time limit.
In order to gain additional insight into the properties of the average heat, we now analyze the asymptotic behavior at high and low temperatures.

\subsubsection{Asymptotic  analysis}

\begin{figure}[!h]
\centering
\includegraphics[width=0.4\textwidth]{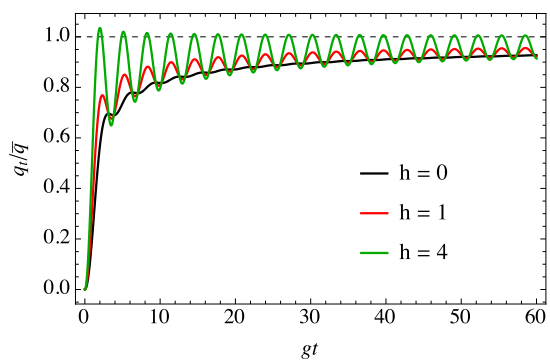}
\caption{\label{fig:aveQ_highT} Average heat $q_t/\bar{q}$ as a function of time, computed using Eq.~(\ref{q_highT}) with $J = 1$, $T_1 = 10$, $T_2 = 12$ and several values of $h$.}
\end{figure}

At high temperatures, $T\gg \lambda_k$, we may expand the occupation numbers $n_k^1$ and $n_k^2$ in Eq.~(\ref{nk}) in a power series in $\lambda_k/T$. 
This allows us to write the average heat density~(\ref{qt}) as 
\begin{equation}\label{q_highT}
q_t \simeq \frac{\Delta \beta}{8} [h^2+2J^2-I(gt)]
\end{equation}
where $I(gt)$ may be expressed in terms of Bessel functions $J_n(x)$ as 
\begin{equation}
I(u)= 2 J^2 J_1(u) \sin u+ (2J^2+h^2) J_0(u) \cos u
\end{equation}
This result shows that at high temperatures the average heat relaxes algebraically to the long-time density 
\[
\bar{q} \simeq  \frac{\Delta \beta}{8} [h^2+2J^2]
\]
with a leading behavior $(gt)^{-1/2}$. 
Eq.~(\ref{q_highT}) is illustrated in Fig.~\ref{fig:aveQ_highT} for different values of $h$. Here and henceforth, in all numerical analyzes, we set $J=1$, thereby redefining the energy scale. 
As can be seen, the heat gradually tends to its long-time value, either monotonically, or in an oscillatory fashion. 
The magnitude of the oscillations increases with increasing field $h$. 
Note that since we are at high temperatures, the quantum phase transition at $h = 2J = 2$ is imperceptible. 

Next we turn to low temperatures, where the effects of the quantum phase transitions  become manifest. 
In this case we will restrict our analysis to the case where $T_1 = T$ and $T_2 = T+\Delta T$. 
The average heat is given in Eq.~(\ref{Qave}). Below the critical point, when $h < 2 J$, we change variables to $u = \lambda/T = (h-2J\cos k)/T$ and exploit the fact that $T$ is small to write Eq.~(\ref{Qave}) as 
\[
q_t  \simeq \frac{T \Delta T}{\pi\sqrt{(2J)^2-h^2}} \int\limits_{(h-2J)/T}^{(h+2J)/T} \ud u \frac{u^2 e^u}{(e^u+1)^2} \sin^2(\alpha_t + \gamma_t u)
\]
where
\[
\alpha_t=gt \sin^2 k_F=gt \left[1-\left(\frac{h}{2J}\right)^2\right],\qquad  \gamma_t=gt \frac{2h T}{ (2J)^2}
\]
($k_F$ is the Fermi momentum of the system).
Extending the limits of integration to $\pm \infty$ allows us to obtain the closed formula:
\begin{IEEEeqnarray}{rCl}
\label{q_below} q_t &=& \frac{\pi T \Delta T}{6 \sqrt{(2J)^2-h^2}} \Bigg\{ 1 - \frac{3 \cos(2\alpha_t)}{\sinh(2\pi \gamma_t)^3} \Big[\sinh(4\pi \gamma_t)\nonumber \\[0.2cm]
&& -\pi \gamma_t (3+ \cosh(4 \pi\gamma_t)) \Big]\Bigg\}
\end{IEEEeqnarray}
which is valid for $h < 2J$. 
When $gt\to\infty$ we obtain 
\begin{equation}\label{longtime_below}
\bar{q} =  \frac{\pi T \Delta T}{6 \sqrt{(2J)^2-h^2}}
\end{equation}

\begin{figure}[!h]
\centering
\includegraphics[width=0.22\textwidth]{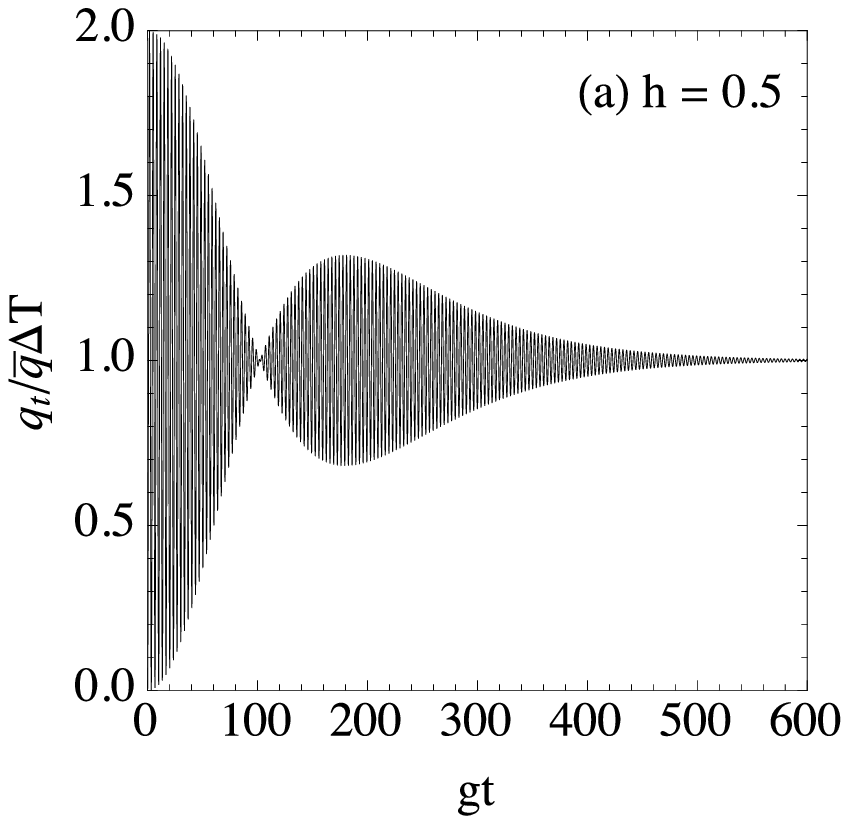}\quad
\includegraphics[width=0.22\textwidth]{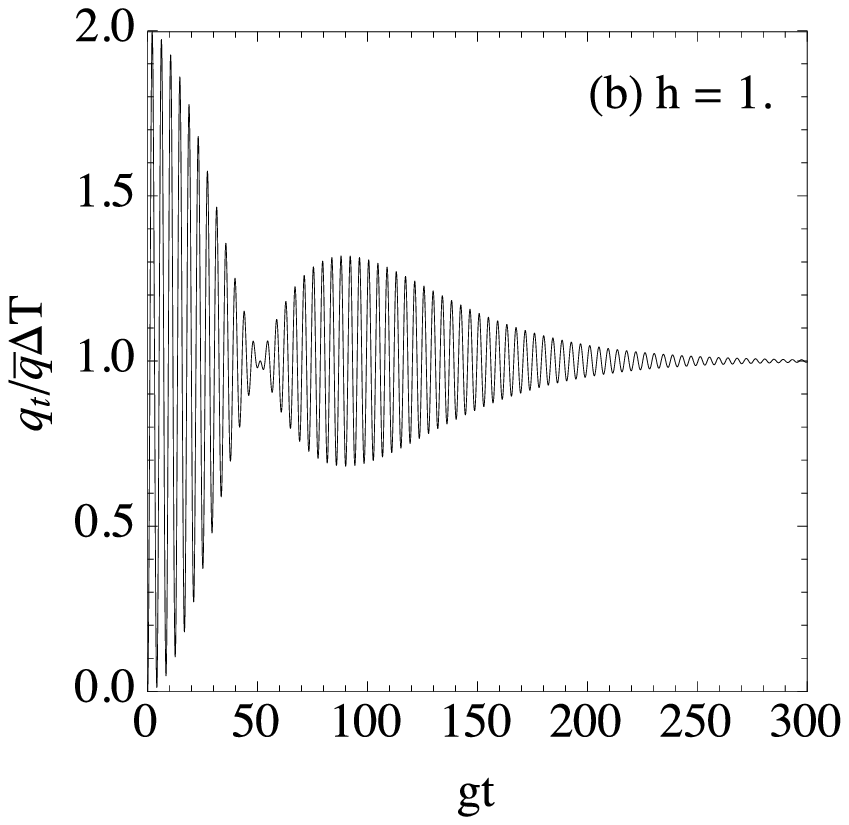}\\
\includegraphics[width=0.22\textwidth]{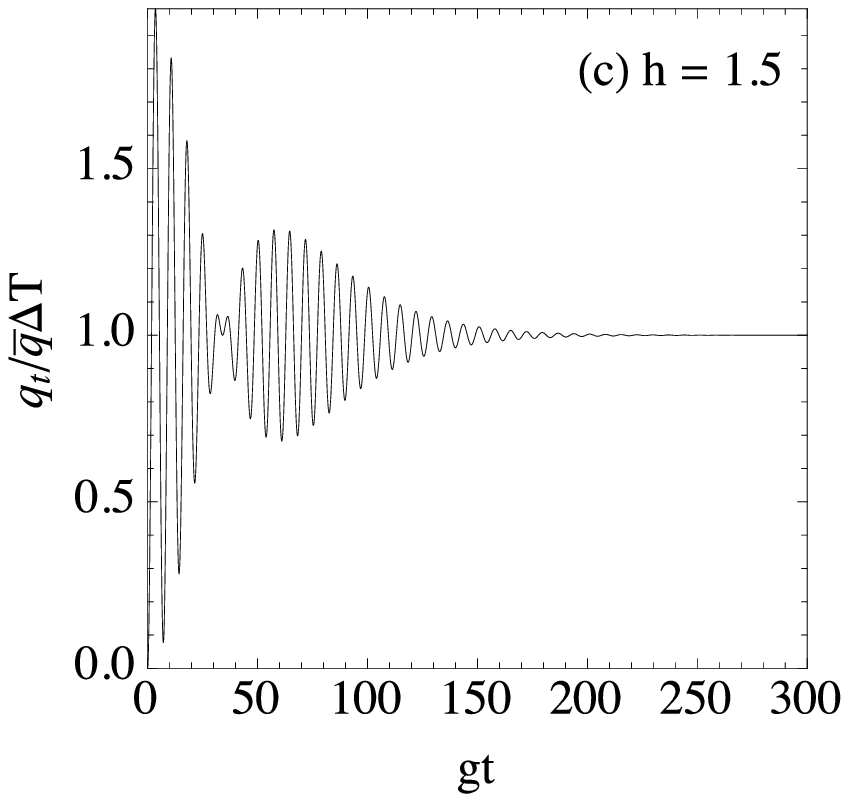}\quad
\includegraphics[width=0.22\textwidth]{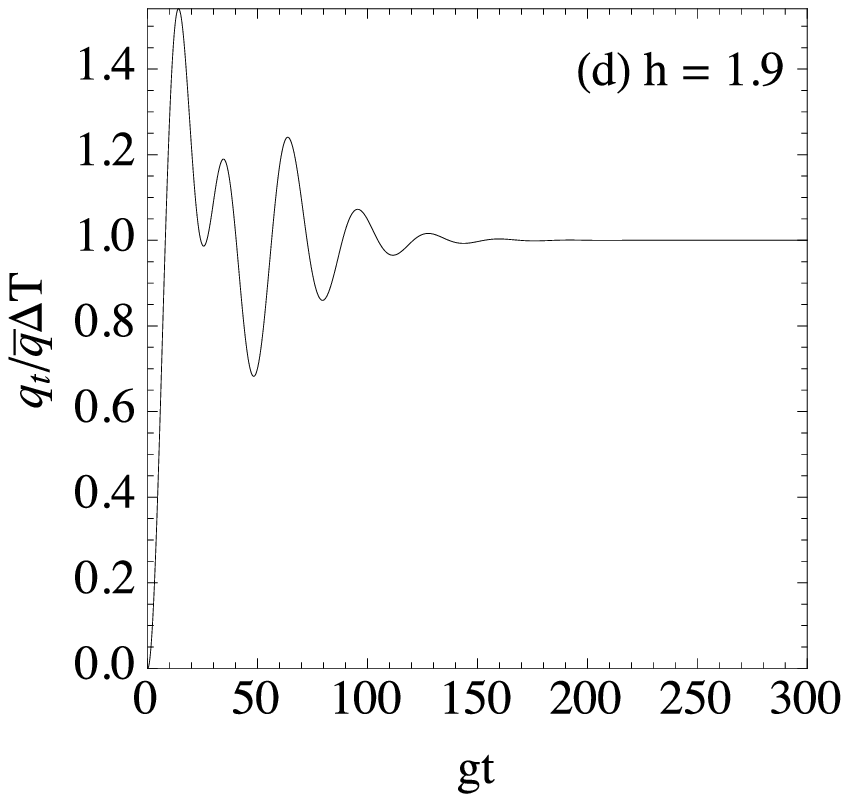}
\caption{\label{fig:aveQ_lowT_below} Average heat in the thermodynamic limit at low temperatures and below the quantum critical point, computed using Eq.~(\ref{q_below}) with $T = 0.01$ and different values of $h$. 
}
\end{figure}

Eq.~(\ref{q_below}) shows the existence of two time scales. 
One is related to the slow relaxation toward equilibrium and is governed by the factor $\gamma_t$, whereas the other is responsible for an oscillatory behavior  governed by $\alpha_t$.  These results are illustrated in Fig.~\ref{fig:aveQ_lowT_below} for different values of $h$. 
As can be seen, the period of the oscillations become longer as $h$ approaches $2J$. At the critical point we have $\alpha_t \to 0$ and the oscillations disappear entirely. 

Next we turn to the low temperature behavior above the critical point, $h > 2J$. 
To obtain an approximate formula in this case we adopt a different strategy and exploit the fact that in this case only excitations close to the ground state $k = 0$ will contribute. We may then expand $\cos k$ and $\sin k$ in a power series in $k$ and extend the upper limit of integration of Eq.~(\ref{Qave}) to $+\infty$. Letting $x = \sqrt{\frac{J}{T}} k$, we then obtain 
\[
q_t \simeq \frac{\Delta T}{\pi} \sqrt{\frac{T}{J}}  e^{-\eta} \int\limits_0^\infty \ud x (\eta+x^2)^2 e^{-x^2}\sin^2(\theta_t x^2/2)
\]
where 
\[
\eta = \frac{h-2J}{T},\qquad \theta_t = \frac{2J\gamma_t}{h} = \frac{ 2 g t T}{J}
\]
The resulting integral may be split into a series of Gaussian integrals, leading to the result
\begin{IEEEeqnarray}{rCl}
\label{q_above} q_t &=& \frac{\Delta T }{16 \sqrt{\pi} }\sqrt{\frac{T}{J}} e^{-\eta} \Bigg\{3+ 4 \eta + 4 \eta^2 
- 3 \frac{\cos(\frac{5}{2} \tan^{-1}\theta_t)}{(1+\theta_t^2)^{5/4}} 
\nonumber  \\[0.2cm]
&&- 4 \eta \frac{\cos(\frac{3}{2} \tan^{-1}\theta_t)}{(1+\theta_t^2)^{3/4}}
-4 \eta^2  \frac{\cos(\frac{1}{2} \tan^{-1}\theta_t)}{(1+\theta_t^2)^{1/4}} 
\Bigg\}
\end{IEEEeqnarray}
The exchanged heat is therefore found to decay exponentially with $\eta = (h-2J)/T$. This is a direct consequence of the quantum phase transition: the opening of a gap between the ground state and first excited state of each chain dramatically reduces the probability of exchanging excitations at low temperatures. 
The trigonometric quantities in Eq.~(\ref{q_above}) may all be written as algebraic combinations of $\theta_t$.
Hence, above the critical points the oscillations of $q_t$ vanish entirely,  giving place to a monotonically increasing behavior, from $q_0 = 0$ toward the long-time density
\begin{equation}\label{longtime_above}
\bar{q} = \frac{\Delta T }{16 \sqrt{\pi} }\sqrt{\frac{T}{J}} e^{-\eta} \left(3+ 4 \eta + 4 \eta^2  \right)
\end{equation}

\begin{figure}[!h]
\centering
\includegraphics[width=0.4\textwidth]{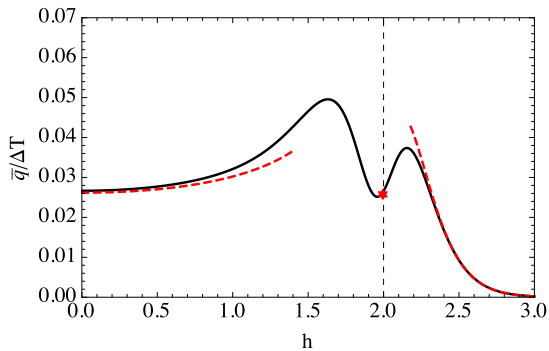}
\caption{\label{fig:lowT_longtime}Long time average heat density as a function of $h$, with $T = 0.1$, computed using Eqs.~(\ref{longtime_below}) and (\ref{longtime_above}) (red lines) compared with the exact numerical solution of Eq.~(\ref{Qave3}) (black circles).
The red dot at $h= h_c = 2$ corresponds to Eq.~(\ref{q_at_hc}).
}
\end{figure}

The long-time heat $\bar{q}$, below [Eq.~(\ref{longtime_below})] and above [Eq.~(\ref{longtime_above})] the quantum critical points are illustrated in Fig.~\ref{fig:lowT_longtime}, which also presents a comparison with the numerical solution of Eq.~(\ref{Qave3}). Below the critical point it increases algebraically with $h$, whereas above the critical point it decreases exponentially. 
Exactly at the critical point, $h = 2J$, a more detailed analysis of $\bar{q}$ is required. In this case we obtain
\begin{IEEEeqnarray}{rCl}
\bar{q} &=& \frac{\Delta T}{\pi} \sqrt{\frac{T}{4J}} \int\limits_0^\infty \ud u \frac{u^4 e^{u^2}}{(e^{u^2}+1)^2} \nonumber\\[0.2cm]
\label{q_at_hc}&=&  \frac{\Delta T}{\pi} \sqrt{\frac{T}{4J}} \frac{3\sqrt{\pi}}{8} \left(1-\frac{1}{\sqrt{2}}\right) \zeta(3/2)
\end{IEEEeqnarray}
where $\zeta(x)$ is the Riemann zeta function.

\subsubsection{Finite-size effects}

Thus far we have focused on the average heat density in the thermodynamic limit. 
We now consider how finite-size effects affect the heat flow.
For finite sizes the average heat may be computed using Eq.~(\ref{Qfin}) or, in the close-to-equilibrium case, 
\begin{equation}\label{Qfin}
\langle Q \rangle_t =\Delta T \sum_k c_k(T) \sin^2(gt\sin^2k)\; .
\end{equation}
which is the analogue of Eq.~(\ref{Qave}) for finite sizes. 

\begin{figure}
\centering
\includegraphics[width=0.43\textwidth]{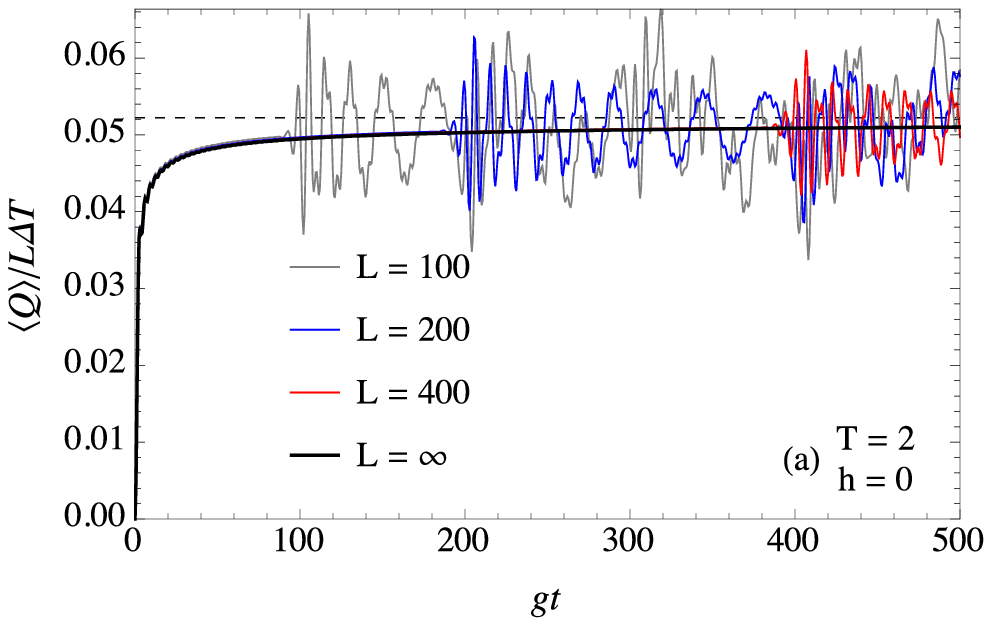}\\
\includegraphics[width=0.43\textwidth]{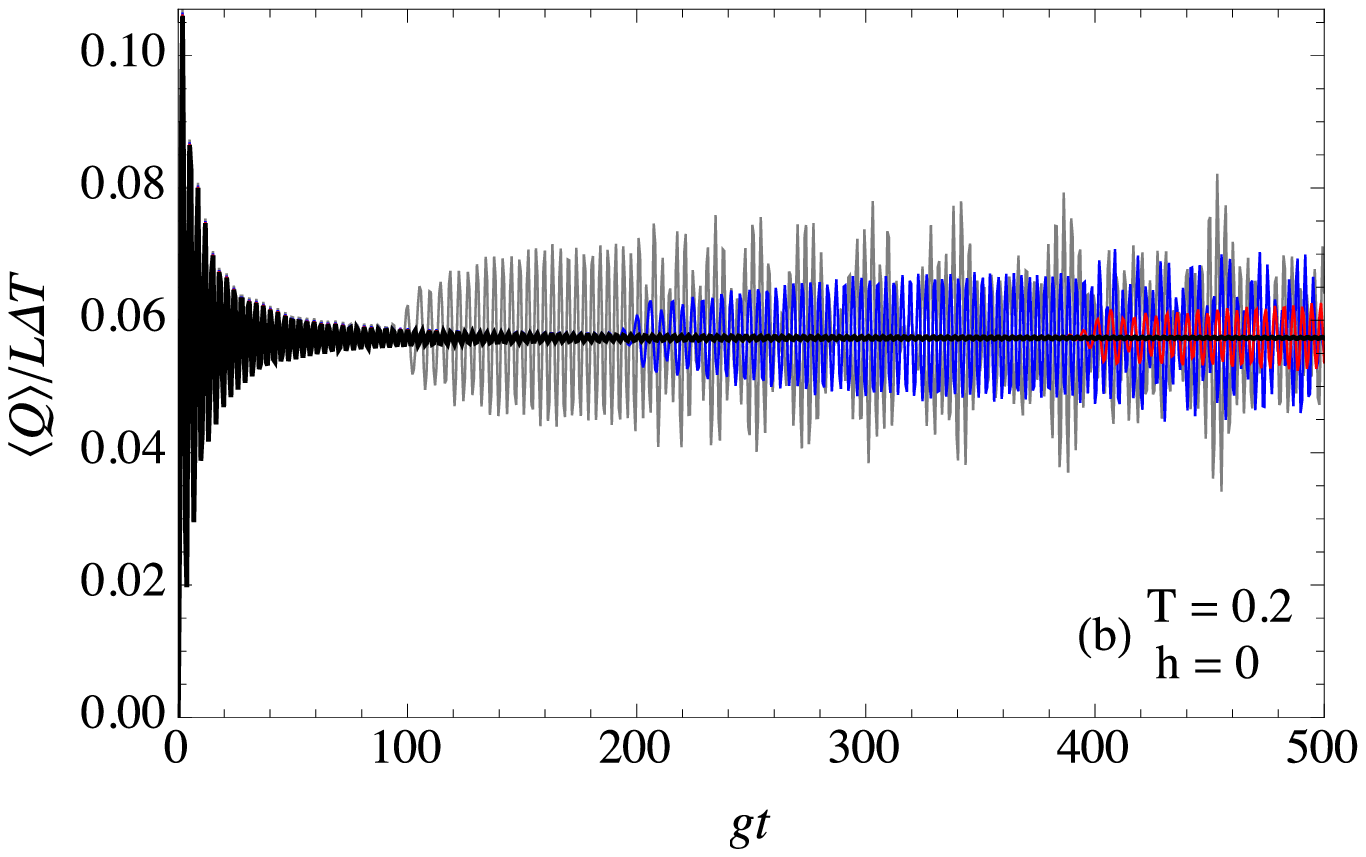}\\
\includegraphics[width=0.43\textwidth]{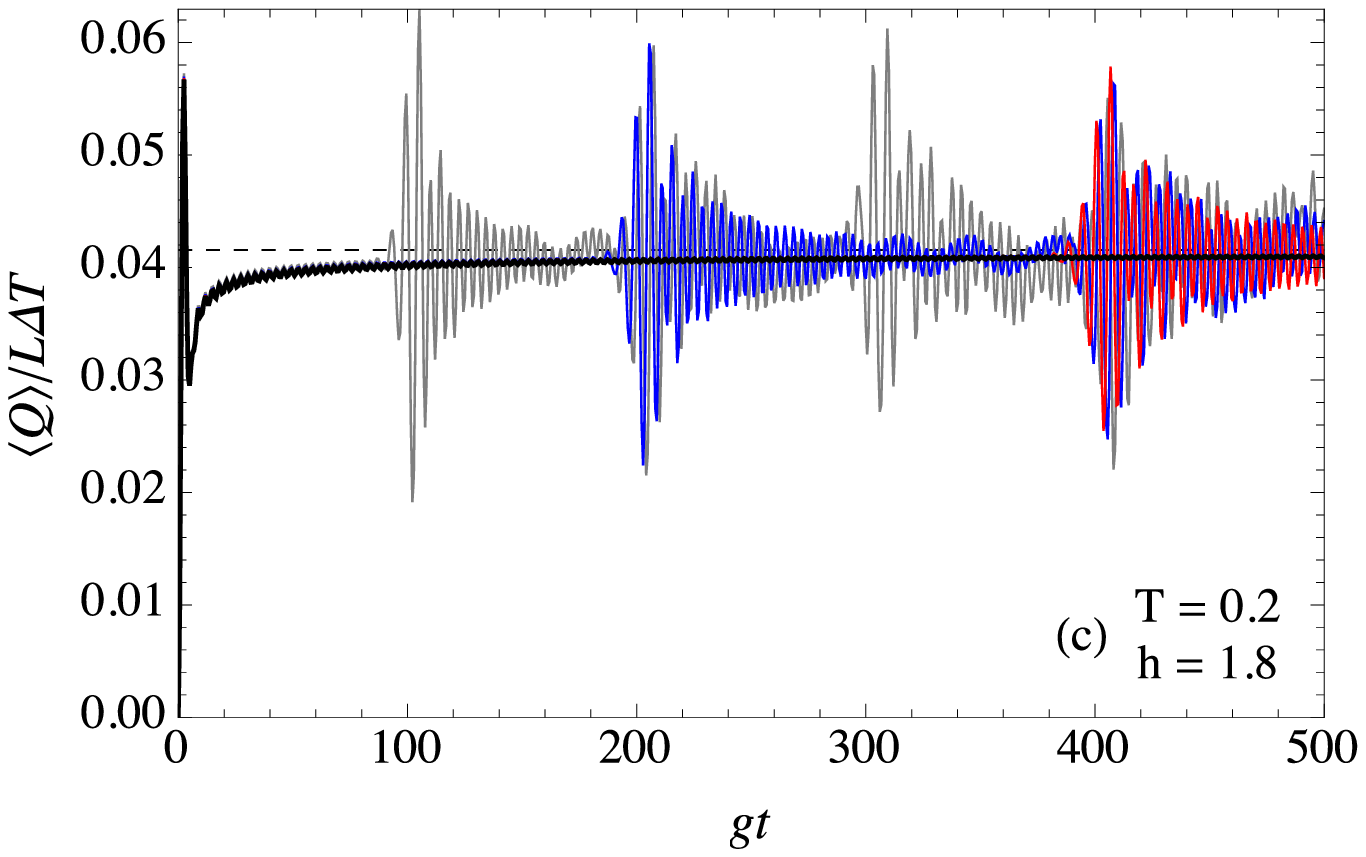}
\caption{\label{fig:heat3}
$\langle Q \rangle/L\Delta T$~vs.~$gt$ for several values of $L$ and different combinations of $h$ and $T$, as shown in each image.
The horizontal dashed lines correspond to Eq.~(\ref{Qave3}).
}
\end{figure}

Results for different sizes and a variety of combinations of  $T$ and $h$ are shown in Fig.~\ref{fig:heat3}. 
As can be seen, the curves for different sizes scale very well up to a certain point in time, after which strong finite-size effects begin to take place, leading to violent oscillations.
The instant where these effects begin to take place scale as $L^2$, in contrast to the time scale $1/g$ which scales proportionally to $L$.  
Hence, as the size of the system increases, these finite size effects are washed away and eventually vanish entirely in the thermodynamic limit. 
It is worth mentioning that these finite size effects are not a consequence of any of the approximations used to get to Eq.~(\ref{Qfin}). 
This is explicitly illustrated in Appendix~\ref{sec:alt}, where these results are compared with the exact diagonalization of the system.

\subsection{\label{sec:PQ}Statistics of heat}

As we have seen, the heat $Q$ is a sum of statistically independent random variables. 
It then follows from the central limit theorem that in the thermodynamic limit the probability distribution $P(\mathcal{Q})$ will be normally distributed. 
The mean of the distribution is $\langle Q \rangle_t$, which was studied in detail in the previous section. 
The variance of the distribution will also be simply the sum of the variances of each individual $Q_k$.  
Proceeding similarly to Eq.~(\ref{Qk_ave}), we then get
\[
\text{var}(Q) = \sum\limits_k \text{var}(Q_k)
\]
where
\[
\text{var}(Q_k) = \lambda_k^2(p_k^+ + p_k^-) - \langle Q_k \rangle_t^2
\]
Hence, the heat distribution in the thermodynamic limit is completely determined. 
For this reason, we will concentrate here on the probability distribution for finite sizes and investigate the role of finite-size effects. 

The distribution $P(\mathcal{Q})$ is the inverse Fourier transform of the characteristic function $F(r)$ in Eq.~(\ref{F_prod}):
\begin{equation}\label{fourier}
P(\mathcal{Q})  = \frac{1}{2\pi}\int\limits_{-\infty}^\infty F(r) e^{-i r \mathcal{Q}} \ud r
\end{equation}
Analytically, it is only possible to carry out this computation in certain  limiting cases. 
But numerically the procedure is trivial for any size desired. 
Here we discuss  3 ways of accomplishing this. 

The most straightforward method to find $P(\mathcal{Q})$ is to note that $Q$ is a sum of the independent discrete random variables $Q_k$ described by the probability density in Eq.~(\ref{pdis}).
Thus, one may  generate random realizations of each $Q_k$ and add them to construct the histogram of $P(\mathcal{Q})$. 

The second method is to expand the product in Eq.~(\ref{Fk}) and compute the Fourier transform of each term independently. 
Since the product will be a sum of complex exponentials in $r$, we will have an equation of the form 
\begin{equation}\label{Finv}
P(\mathcal{Q}) = \sum\limits_\alpha \Gamma(\alpha) \delta(Q - \alpha)
\end{equation}
where the quantities $\alpha$ represent all possible combinations of the form $\pm\lambda_{k_1}\pm\lambda_{k_2}\pm\ldots$, and $\Gamma(\alpha)$ are the corresponding probabilities.
Eq.~(\ref{Finv}) may be used to compute $P(\mathcal{Q})$ for moderate sizes, up to $L = 20$.

Finally, the third way to find $P(\mathcal{Q})$ is to evaluate Eq.~(\ref{fourier}) using the Fast Fourier Transform (FFT) algorithm. 
This method scales as $\mathcal{O}(L \log L)$ and is therefore extremely efficient and applicable to any size desired. 
Unlike the other two methods, however, the FFT is coarse grained and is therefore   recommended only for larger sizes.

\begin{figure}[!t]
\centering
\includegraphics[width=0.22\textwidth]{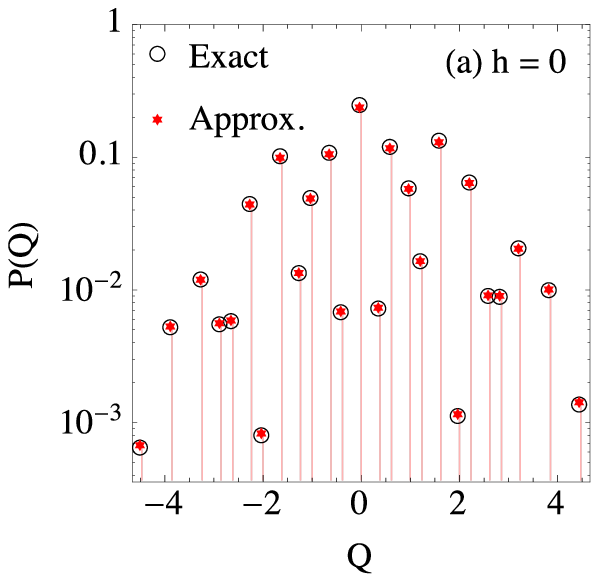}\quad
\includegraphics[width=0.22\textwidth]{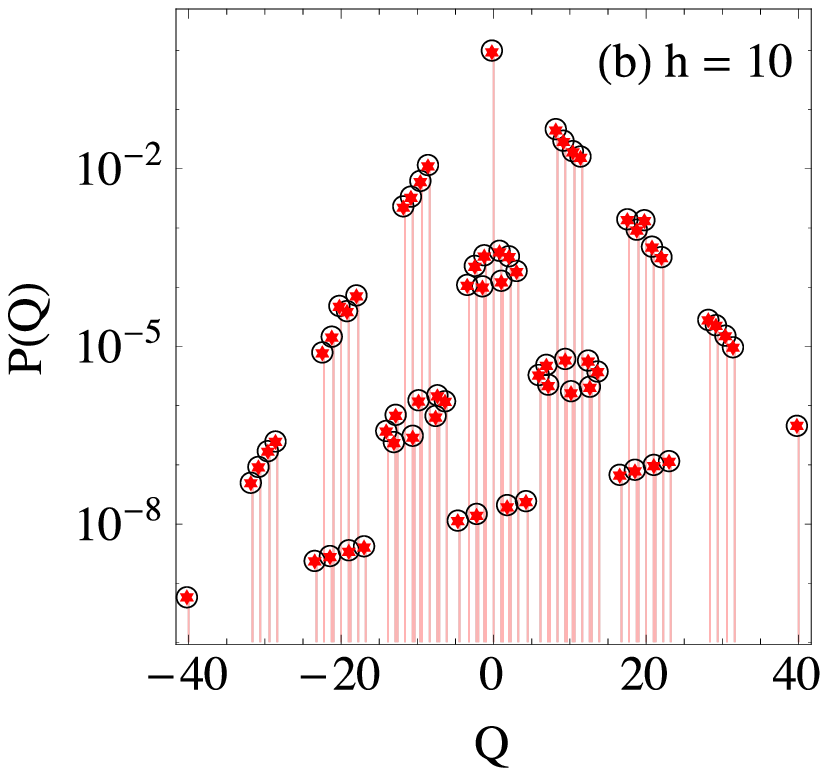}
\caption{\label{fig:PQ_exact}
Comparison between the exact heat distribution, computed from a full diagonalization of the Hamiltonian~(\ref{H}),  and the perturbative solution computed from Eq.~(\ref{Finv}), for $L = 4$, $T_1= 2$, $T_2 = 3$, $g_0 = 0.1$ and $gt = 15$. (a) $h = 0$ and (b) $h = 10$. 
}
\end{figure}

We begin by comparing $P(\mathcal{Q})$ computed from  Eq.~(\ref{Finv}) with the exact result, obtained by diagonalizing  the full Hamiltonian $H$ in Eq.~(\ref{Htot}). This serves as a test of the accuracy of the approximations employed in this paper.
The results are shown in Fig.~\ref{fig:PQ_exact} for $L = 4$ and two different values of $h$. 
In these calculations we must once again make explicit use of the constant $g_0$, which we set at $g_0 = 0.1$ in order to ensure that the coupling is indeed small. 
As can be seen, the agreement between both methods is extremely good, strongly corroborating our main approximation in Eq.~(\ref{Hk}). Further comparisons with the exact solution are made in  Appendix~\ref{sec:alt}.

\begin{figure}
\centering
\includegraphics[width=0.22\textwidth]{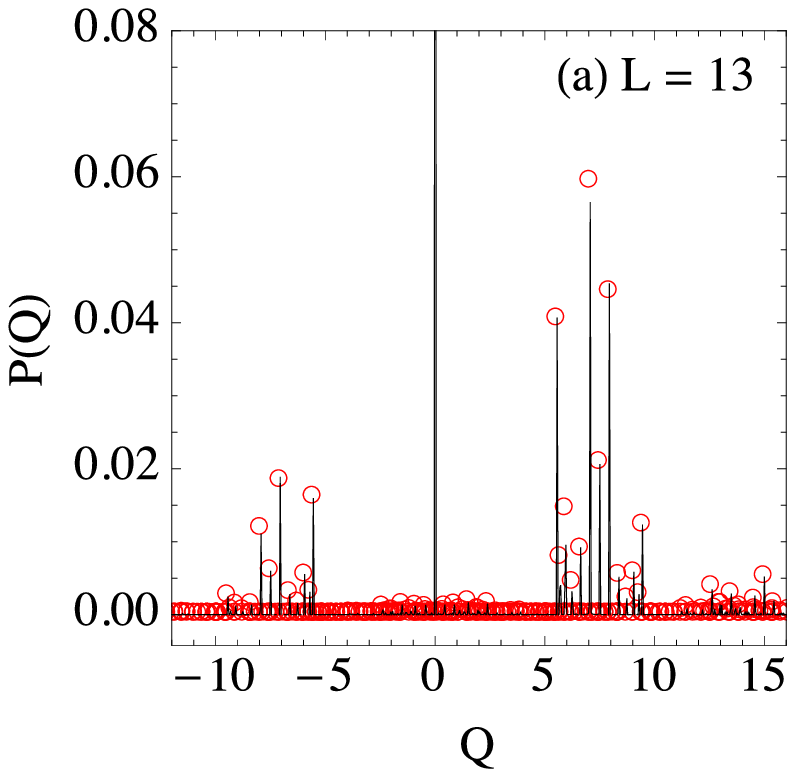}\quad
\includegraphics[width=0.22\textwidth]{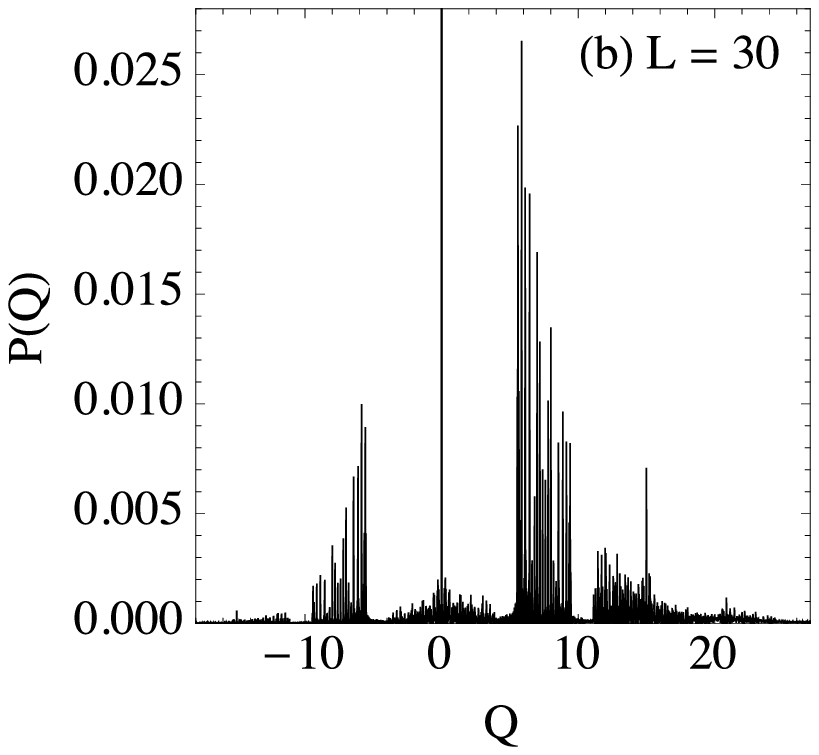}\\
\includegraphics[width=0.22\textwidth]{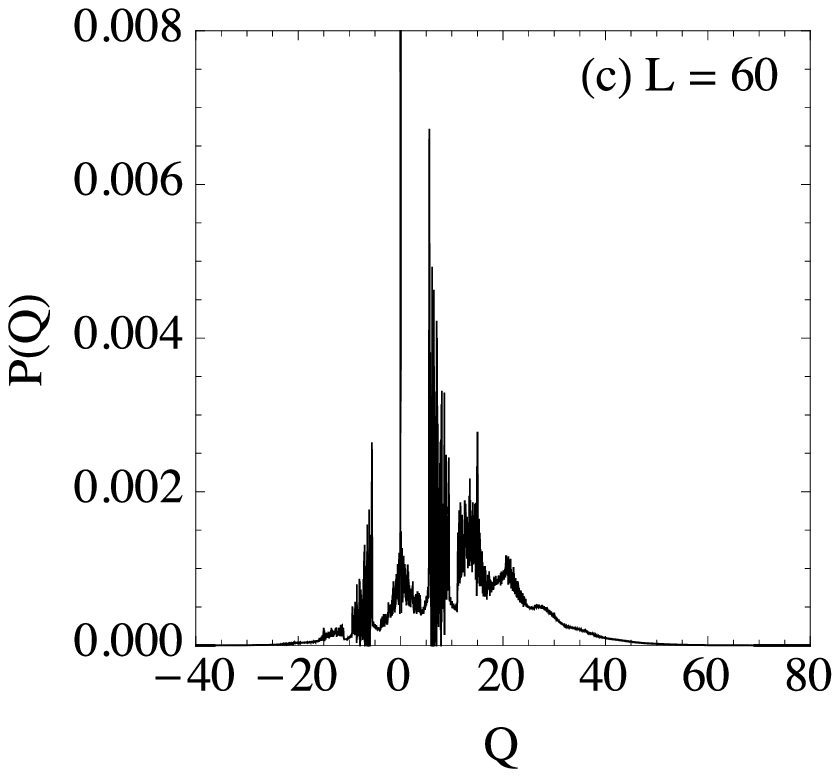}\quad
\includegraphics[width=0.22\textwidth]{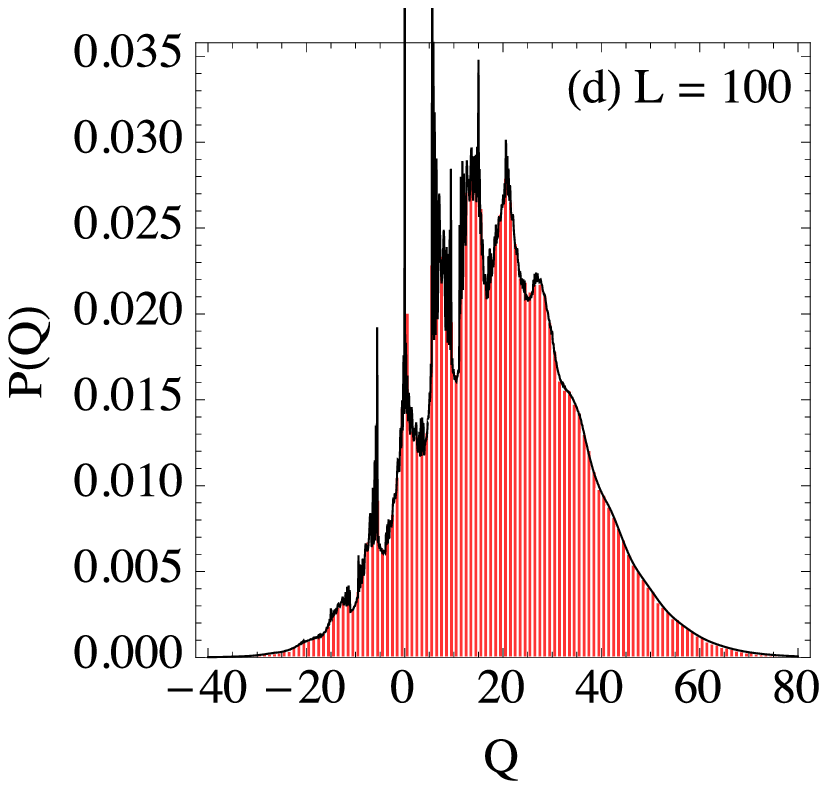}\\
\includegraphics[width=0.22\textwidth]{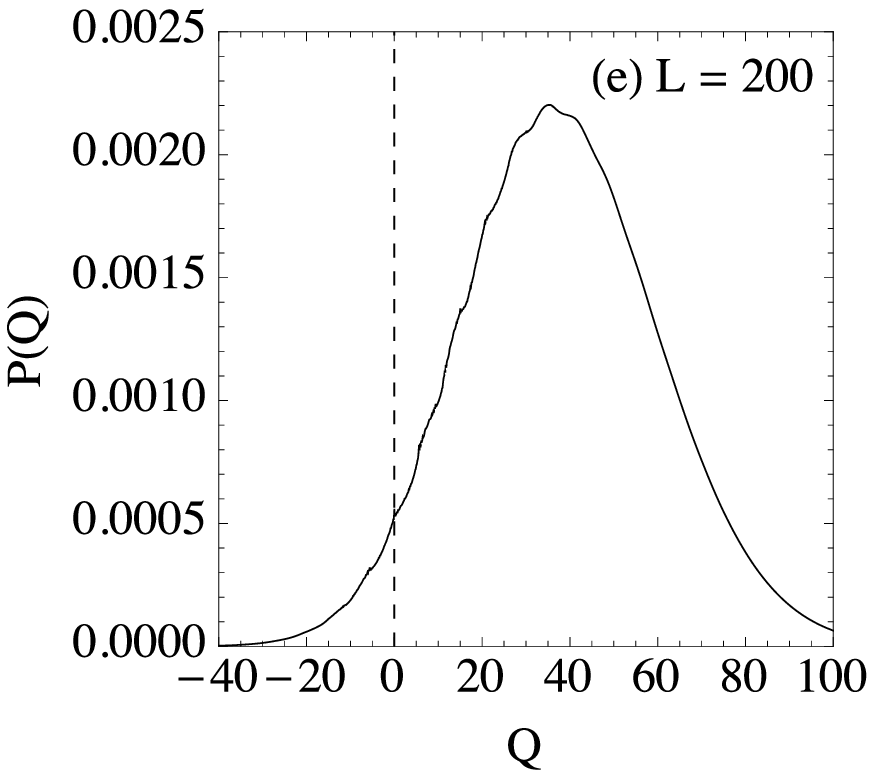}\quad
\includegraphics[width=0.22\textwidth]{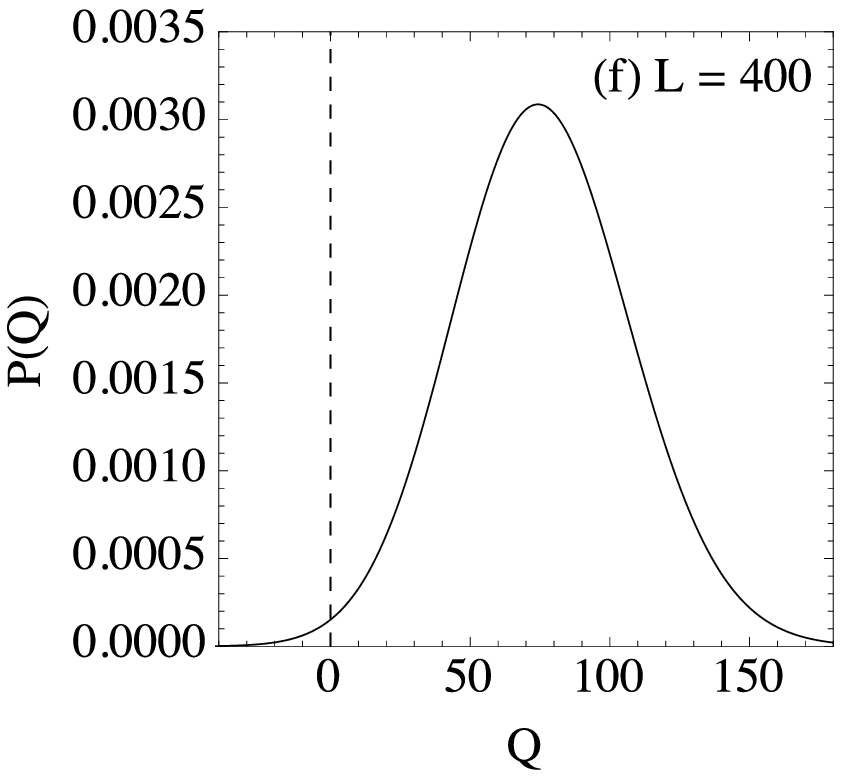}
\caption{\label{fig:PQ}(Color Online)
The distribution of heat $P(\mathcal{Q})$ for  $h = 7.5$, $T = 2$, $\Delta T = 1$ and $gt = 15$, computed using the FFT method. 
In (a) we also present in red a comparison with Eq.~(\ref{Finv}) and in (d) we present a comparison with the random realizations. 
}
\end{figure}

In Fig.~\ref{fig:PQ} we show the distribution $P(\mathcal{Q})$ computed using the FFT method for several sizes. 
In Figs.~\ref{fig:PQ}(a) and (d) we also compare the FFT calculations with the other two methods and, as can be seen, the agreement is found to be very good.  

We see in Fig.~\ref{fig:PQ} that $P(\mathcal{Q})$ gradually transforms from a sharply peaked distribution, to a Gaussian. 
According to the fluctuation theorem in Eq.~(\ref{FT}), the probability of observing a heat flux in the wrong direction should decrease exponentially with the magnitude of $Q$. But since $Q$ is an extensive quantity, this probability should diminish with increasing size. 
This effect can be seen in Fig.~\ref{fig:PQ}, particularly when comparing images (e) and (f).

In the thermodynamic limit the distribution $P(\mathcal{Q})$ will be Gaussian. 
But for finite $L$ the shape of the distribution depends sensibly on the parameters of the system, particularly on the magnetic field $h$. 
Large fields reduce the number of allowed transitions between the energy levels and therefore lead to a more peaked distribution. 
Interestingly, there are certain situations where the shape of the distribution also changes dramatically with time, for a fixed set of parameters. 
This is illustrated in Fig.~\ref{fig:FFTtime}  (see also the corresponding video in the supplemental material). 
During most instants of time the distribution has a simple Gaussian shape. 
However, at certain intervals, it's shape changes completely to a highly peaked structure. 

\begin{figure}
\centering
\includegraphics[width=0.22\textwidth]{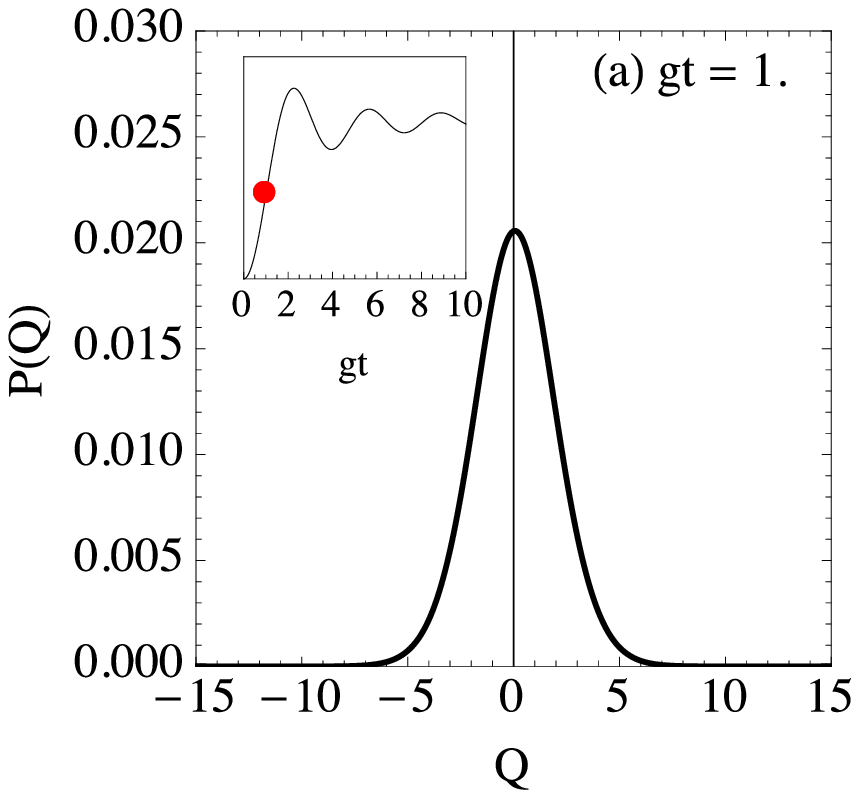}\quad
\includegraphics[width=0.22\textwidth]{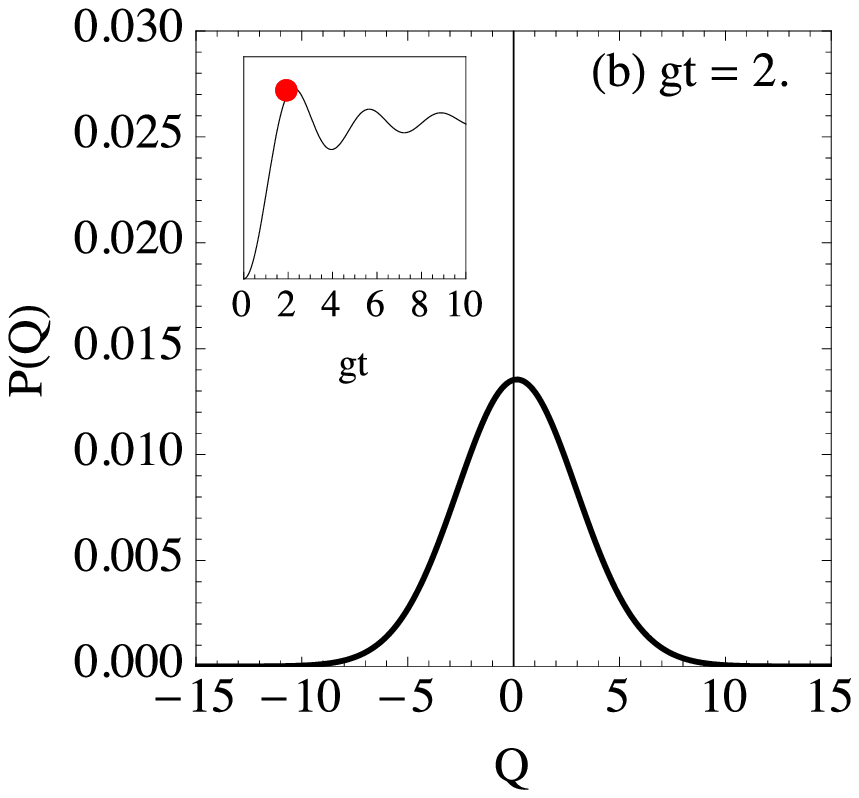}\\
\includegraphics[width=0.22\textwidth]{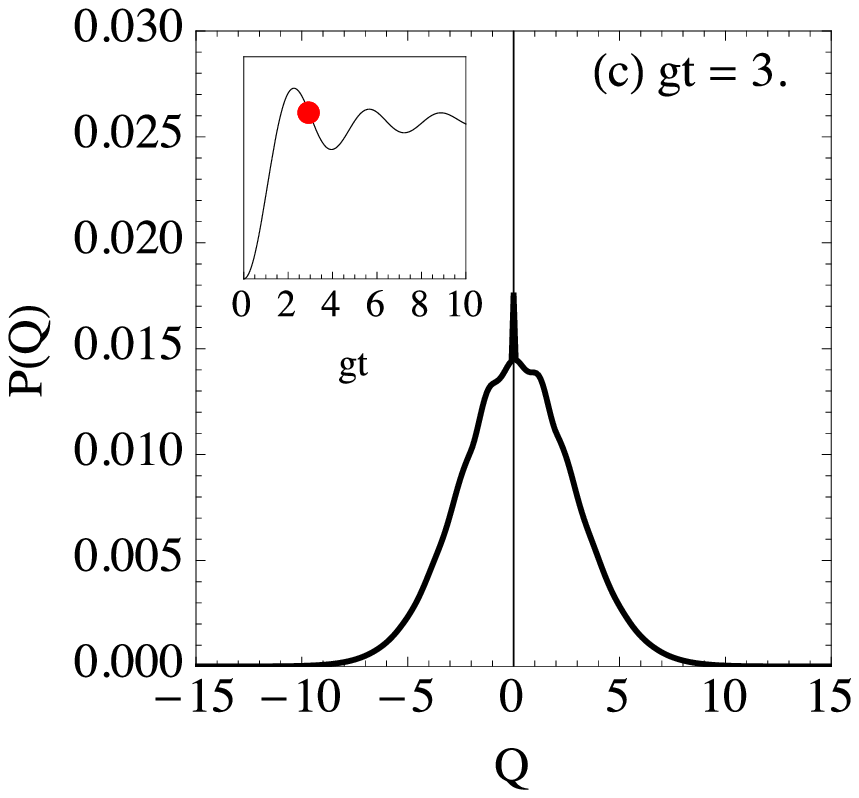}\quad
\includegraphics[width=0.22\textwidth]{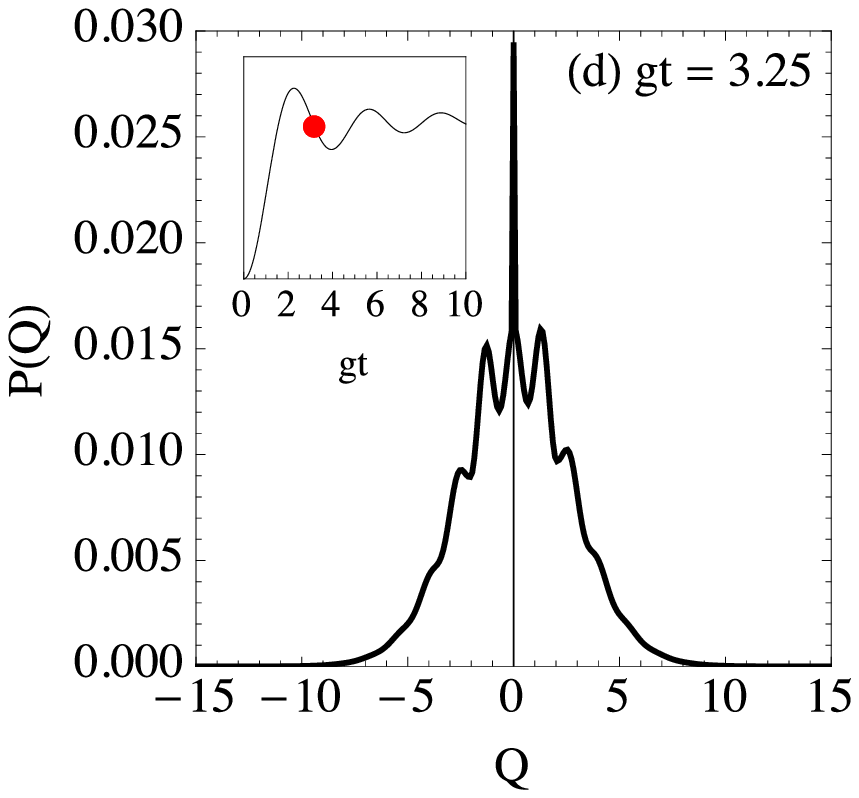}\\
\includegraphics[width=0.22\textwidth]{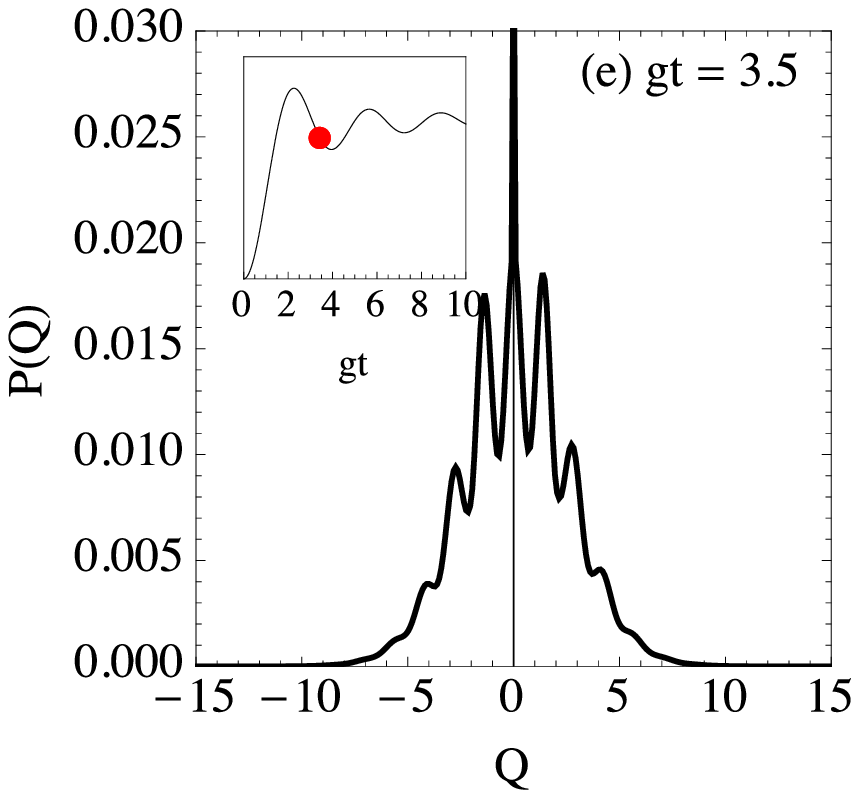}\quad
\includegraphics[width=0.22\textwidth]{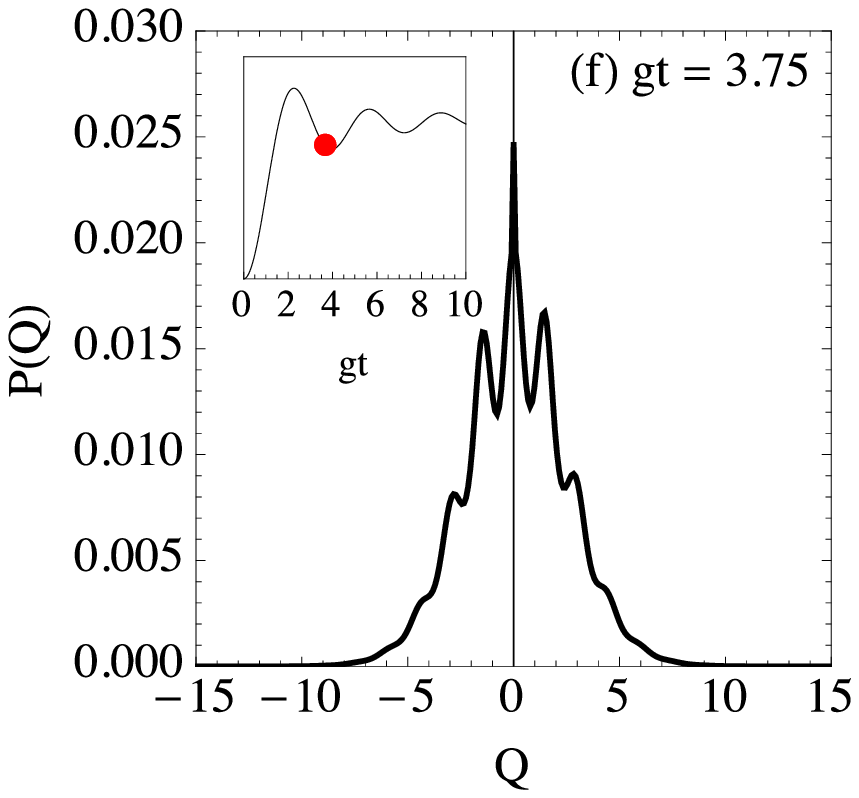}\\
\includegraphics[width=0.22\textwidth]{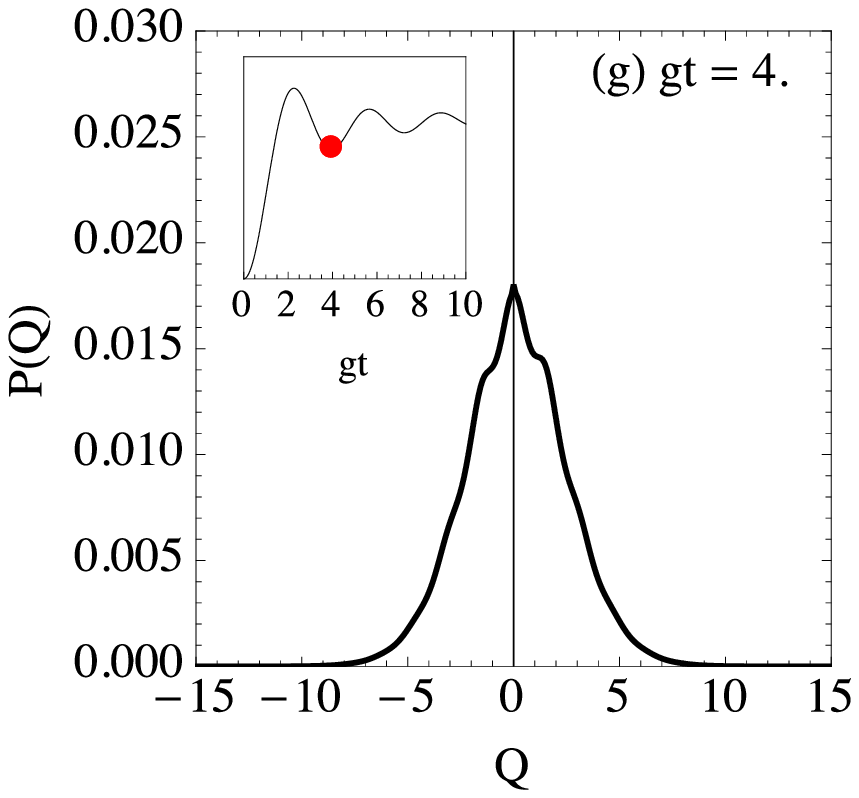}\quad
\includegraphics[width=0.22\textwidth]{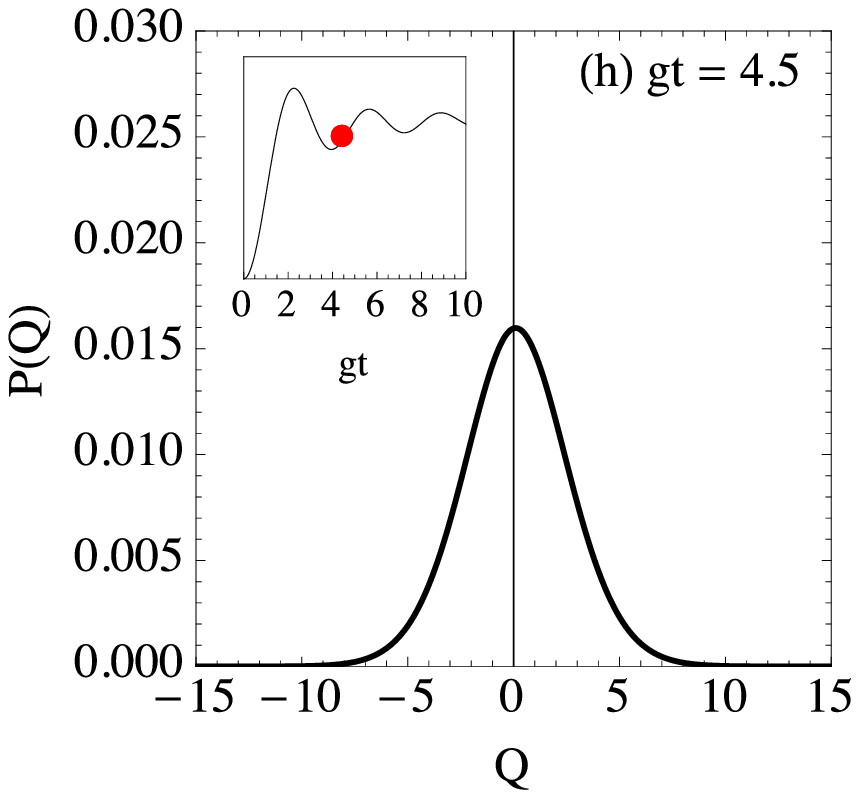}
\caption{\label{fig:FFTtime}
$P(\mathcal{Q})$ at several instants of time, for  $L = 100$, $h = 0$, $T = 0.5$ and $\Delta T = 0.01$. The inset shows the average heat as a function of time and the red dot indicates the corresponding instant.
}
\end{figure}

\begin{figure}
\centering
\includegraphics[width=0.22\textwidth]{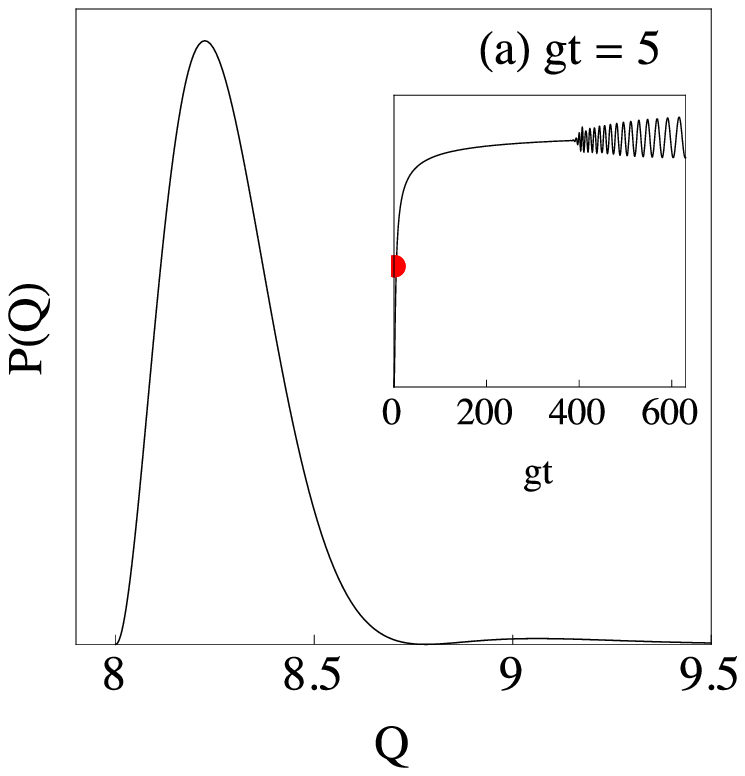}\quad
\includegraphics[width=0.22\textwidth]{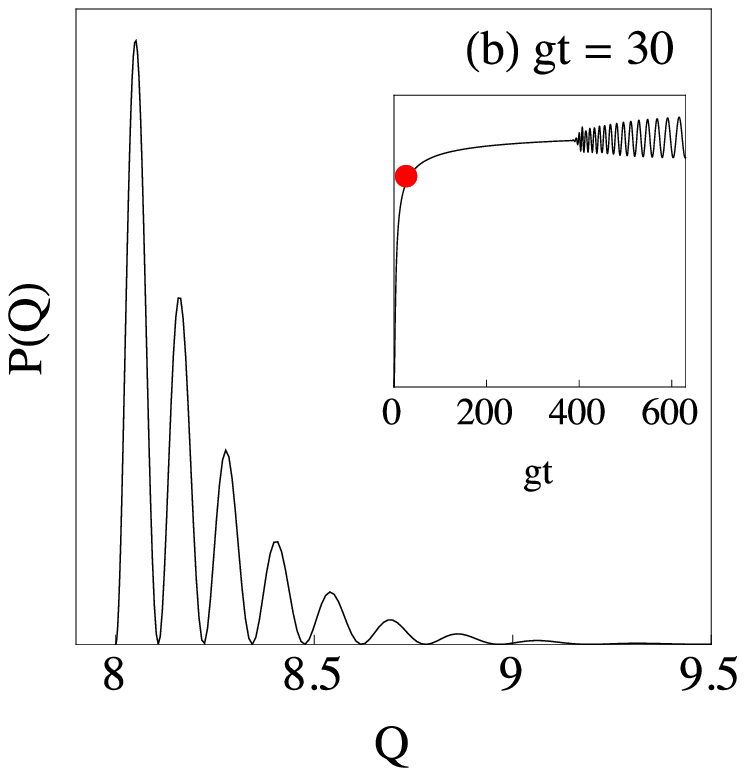}\\
\includegraphics[width=0.22\textwidth]{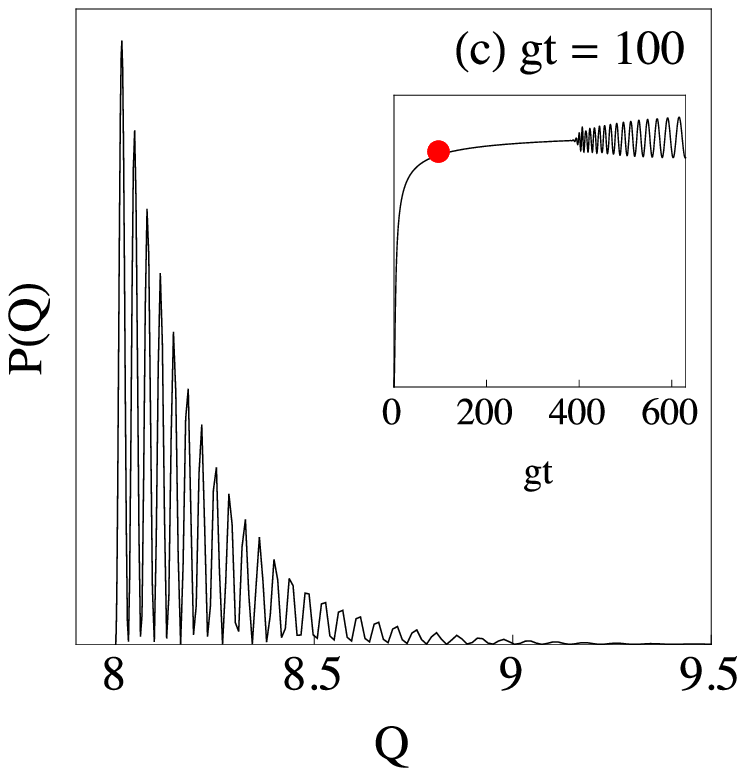}\quad
\includegraphics[width=0.22\textwidth]{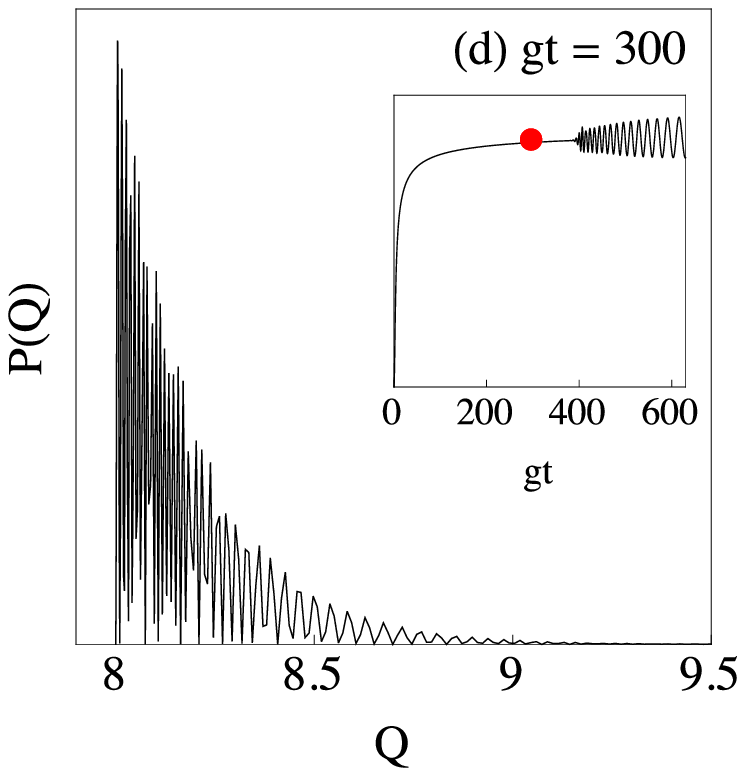}\\
\includegraphics[width=0.22\textwidth]{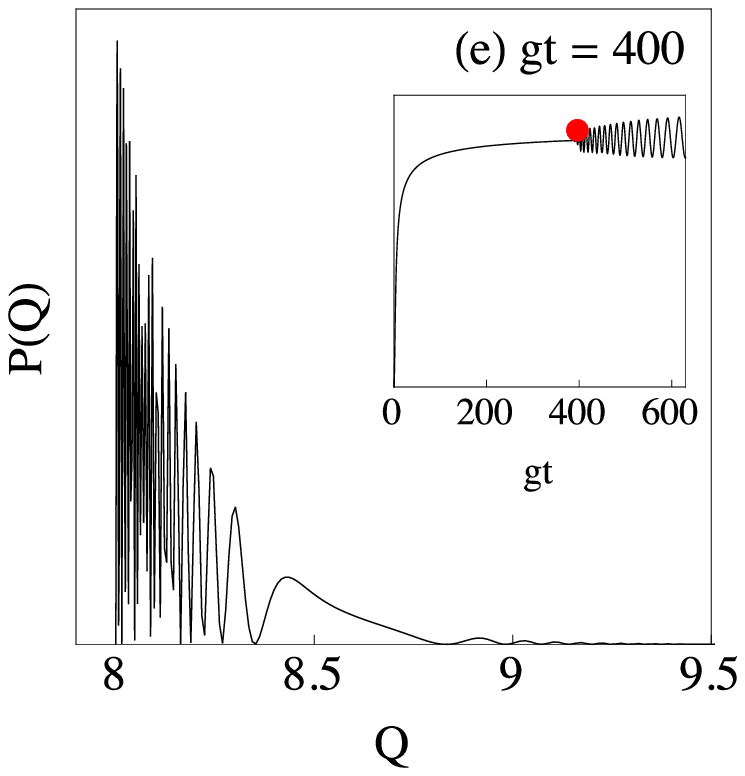}\quad
\includegraphics[width=0.22\textwidth]{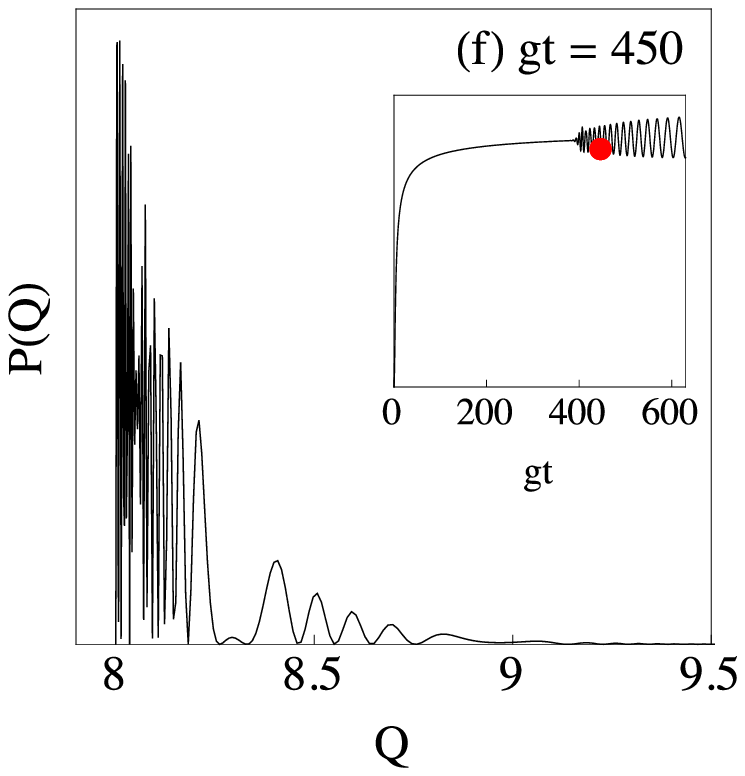}\\
\includegraphics[width=0.22\textwidth]{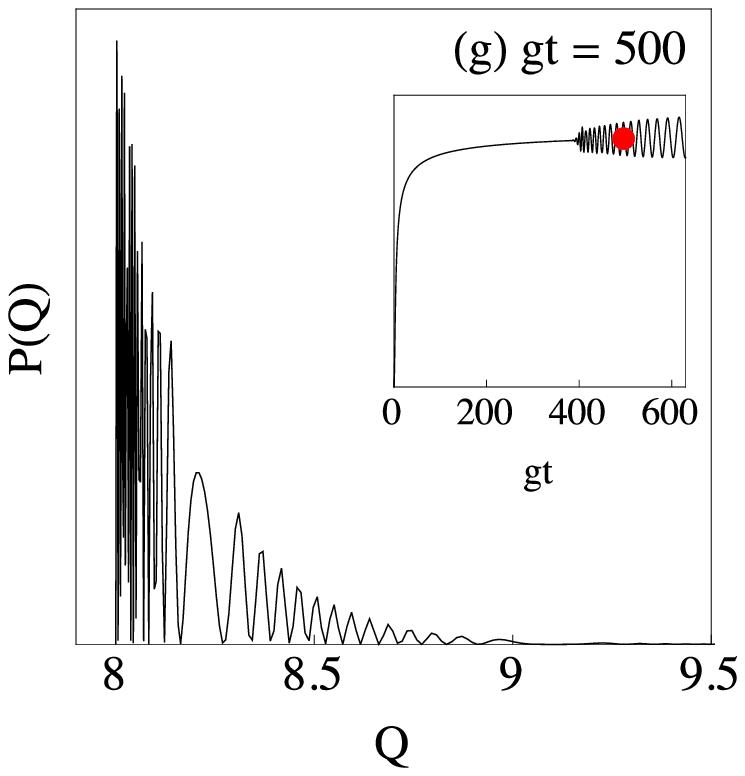}\quad
\includegraphics[width=0.22\textwidth]{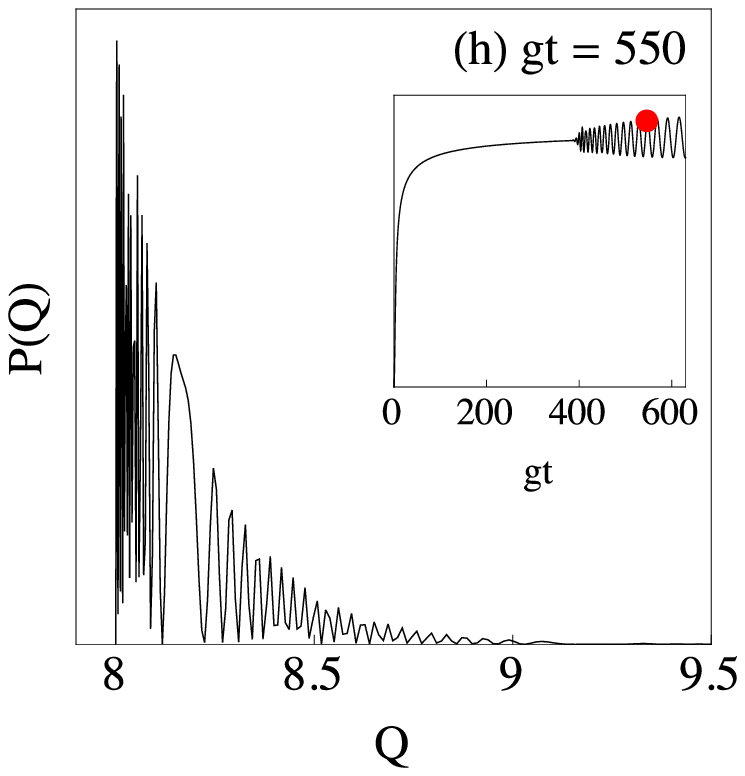}
\caption{\label{fig:FirstBand}
Probabilities of a single excitation, Eq.~(\ref{pkp}), for $L = 400$, $h = 10$ and $T = 0.2$. 
The inset shows the average heat as a function of time. 
}
\end{figure}

As a last topic we turn to the violent finite size oscillations present in the average heat in Fig~\ref{fig:heat3}. 
These oscillations are not readily visible in  $P(\mathcal{Q})$. 
They manifest themselves as small probability fluxes traveling through $P(\mathcal{Q})$ in the form of ripples. 
The best way to visualize them is to consider the particular situation of very low temperatures and high fields. 
In this case it is overwhelmingly more likely that no heat flows at all.
However, there also exists a small but non-zero probability that a single excitation be exchanged between the two chains. 
We will therefore have two bands of probability, one spanning the range $Q = \lambda_k$ and the other spanning the range $Q = - \lambda_k$. 
This can  be seen in Fig.~\ref{fig:PQ} (a) and (b). 
In this case these probabilities will be given by $p_k^+$ and $p_k^-$ in Eqs.~(\ref{pkp}) and (\ref{pkm}) respectively. 

In Fig.~\ref{fig:FirstBand} we plot the probabilities $p_k^+$ for several instants of time when $L = 400$ (see also the accompanying video in the supplemental material). 
As can be seen, as time progresses, $P(\mathcal{Q})$ becomes increasingly more oscillatory, with an accumulation of probability in the left. 
However, as the unstable region approaches, the shape of the distribution changes from a well ordered oscillatory function to a disordered shape. 
Finally, when the violent oscillations start, certain gaps open in the probability distribution, which then propagate from one region to the other in the form of ripples.

%
%
%
%

\section{\label{sec:conc}Conclusions}

We considered the statistics of the heat exchanged between two quantum spin chains of arbitrary size. 
These problems are usually quite complicated to study analytically, even for systems which may be diagonalized exactly. 
The reason is that the distribution of heat requires knowledge of all possible transitions between the energy levels of a many-particle system. 
Consequently, much of the effort in this area has been focused on small systems \cite{Akagawa2009,Gomez-Marin2006,Gomez-Solano2011}. 
The case of many-body systems was studied by Corbieri \emph{et. al.} \cite{Corberi2013} for ferromagnets using the time-dependent Landau-Ginzburg formalism. 
Classical chains of oscillators were studied by Lahiri and Jayannavar \cite{Lahiri}, whereas  quantum oscillator chains were considered by Agarwalla2014  \emph{et. al.} \cite{Agarwalla2014}.

The purpose of this paper was to introduce a framework to allow the investigation of systems of arbitrary size by naturally imposing the physical condition of a weak-coupling between the two systems. 
Such a condition is very reasonable from an experimental point of view and allows for heat to be properly defined.
We use the definition where work is interpreted as changing a parameter in the Hamiltonian. 
All other changes in the energy of the system must be attributed to heat. 
In strong-coupling this distinction is not clear, but for weak-coupling it is. 

When weak-coupling is imposed, the problem factors into $L$ independent systems. 
Consequently, the characteristic function factors as a product, which allows one to compute all expectation values of interest and the probability of heat itself. 
Using this framework we have shown how the distribution of heat and the average heat change as one passes from small to large sizes. 
It was found that finite size effects play a vital role in the problem, manifesting themselves as highly delta-peaked structures for the distribution and violent oscillations for the average heat. 
Conversely, in the thermodynamic limit one recovers a Gaussian distribution for the heat values together with a well-behaved time dependence for the average heat. 
In our view, the methodology employed here may also be readily extended to other many-body problems of similar structure (see Appendix~\ref{sec:work}). 
Therefore, it may be of use to scientists working with quantum thermodynamics and non-equilibrium statistical mechanics. 

\begin{acknowledgements} 

We acknowledge the S\~ao Paulo Research Foundation (FAPESP) for the financial support.
Moreover, I am thankful to Pedro Gomes, Mario Jos\'e de Oliveira, Roberto Serra, Fernando Semi\~ao, Mauro Paternostro, John Goold and Eric Lutz, for fruitful discussions. 

\end{acknowledgements}

%
%
%
%

\appendix
\section{\label{sec:alt}Numerically exact calculation of expectation values}

\begin{figure}
\centering
\includegraphics[width=0.42\textwidth]{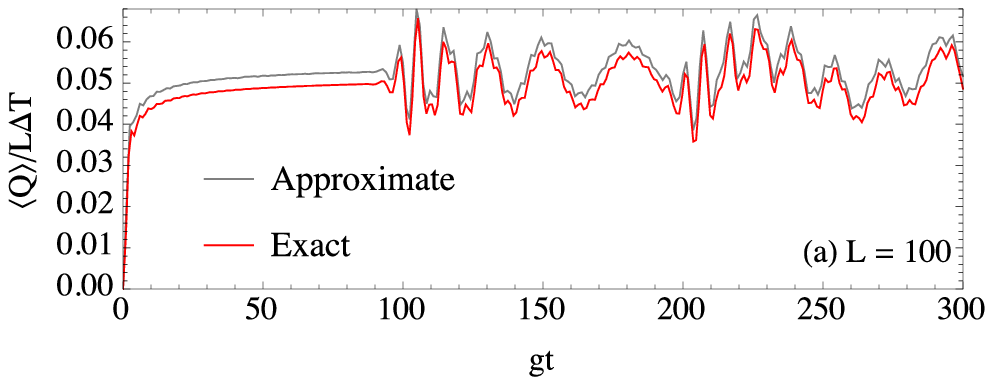}\\
\includegraphics[width=0.42\textwidth]{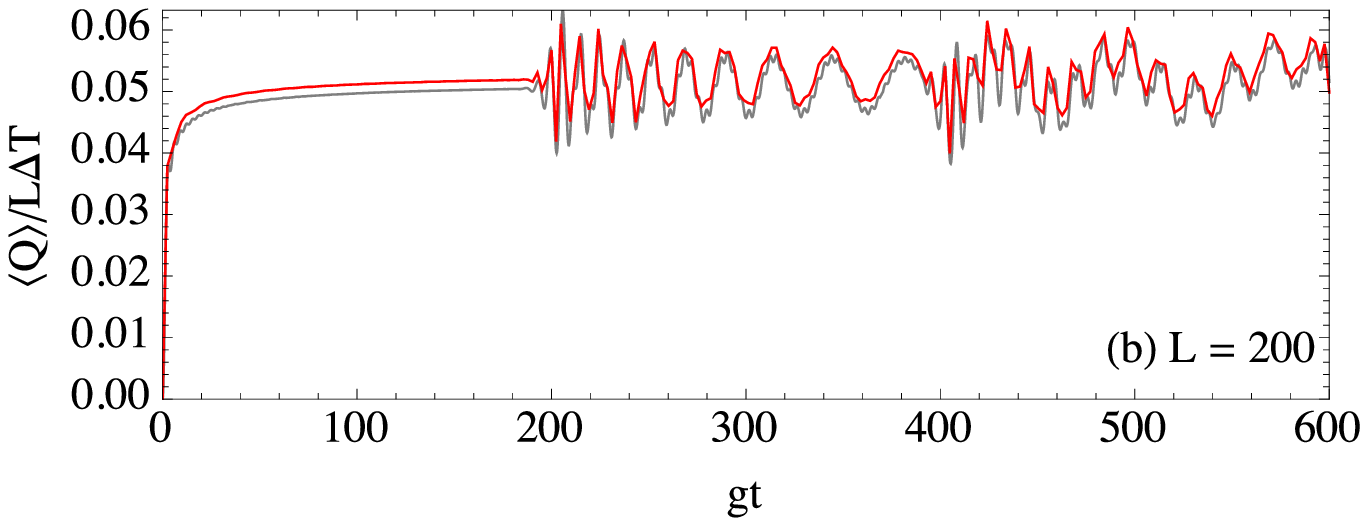}\\
\includegraphics[width=0.42\textwidth]{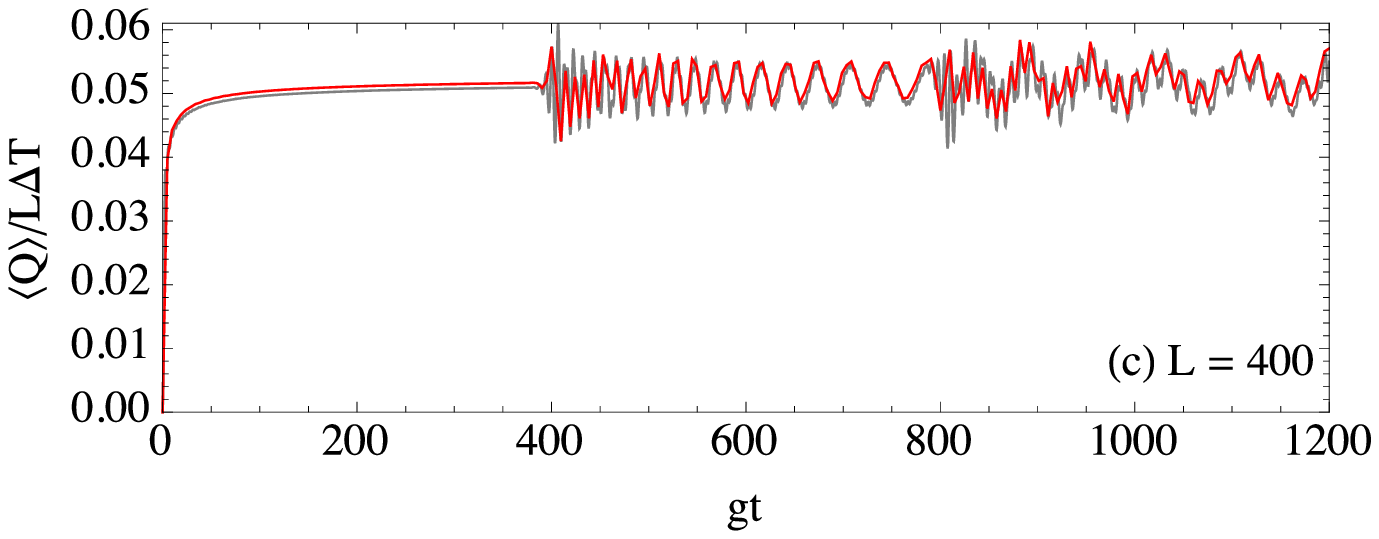}
\caption{\label{fig:NumExact1}
Comparison between the approximate solution for the average heat, Eq.~(\ref{Qave}), and the numerically exact solution computed using Eq.~(\ref{theta_sol}) for $h = 0$, $T = 2$, $g_0 = 0.05$ and different values of $L$. 
}
\end{figure}

\begin{figure}
\centering
\includegraphics[width=0.42\textwidth]{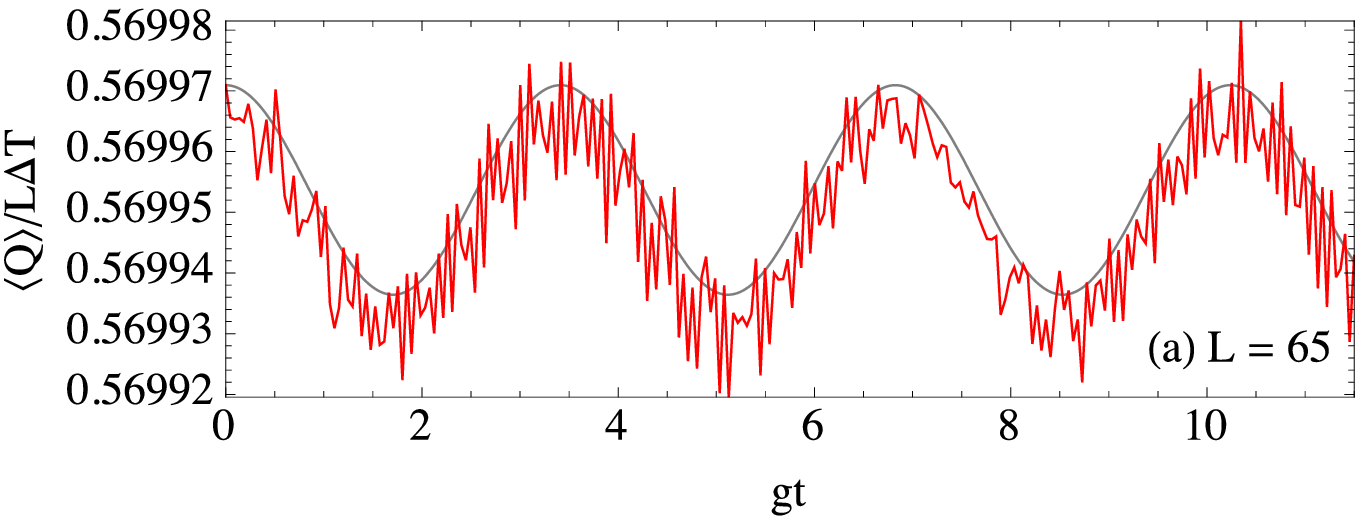}\\
\includegraphics[width=0.42\textwidth]{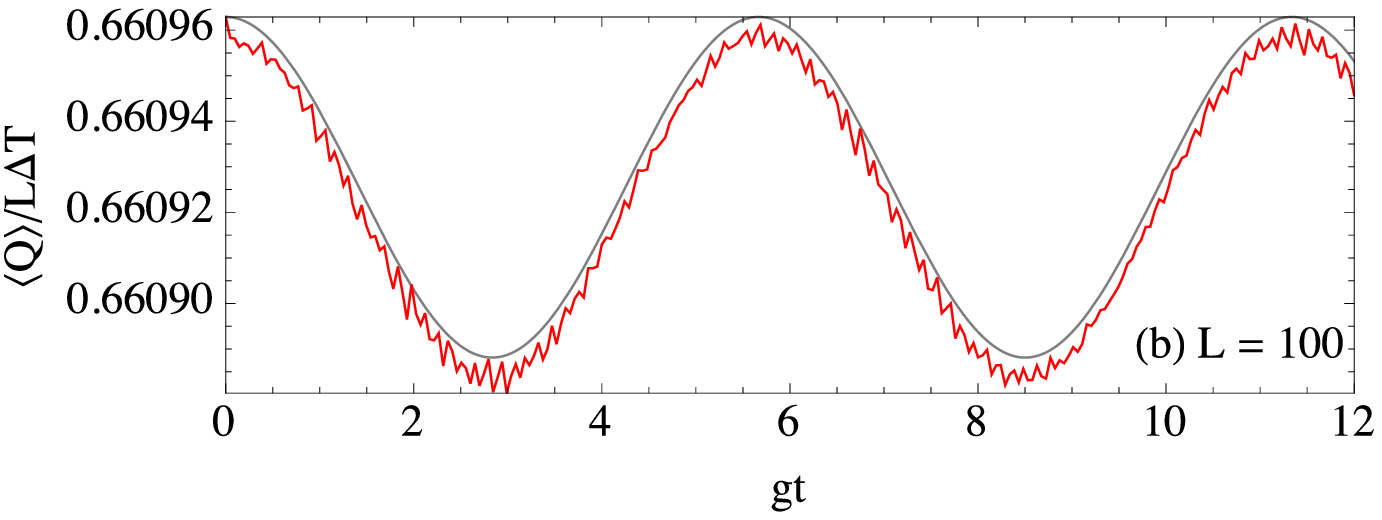}\\
\includegraphics[width=0.42\textwidth]{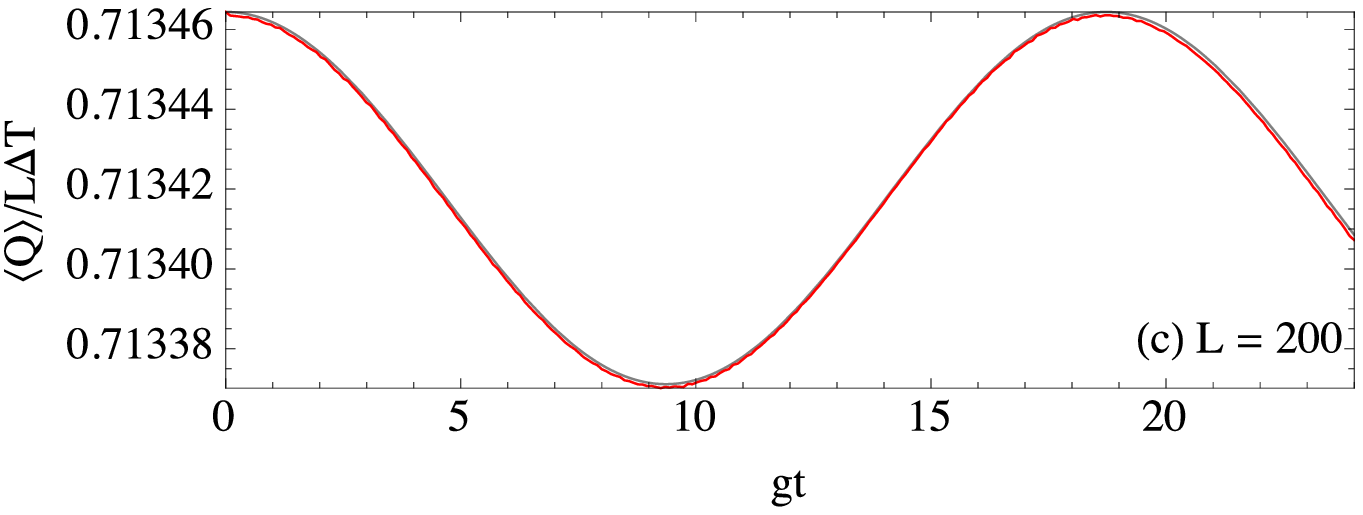}
\caption{\label{fig:NumExact2}
Comparison between Eq.~(\ref{aa}) for an arbitrary $k = 27\pi/(L+1)$, and the numerically exact solution computed using Eq.~(\ref{theta_sol}), for the same configuration of Fig.~\ref{fig:NumExact1}.
}
\end{figure}

In this appendix we discuss a numerically exact method to compute expectation values, which may be used to test the accuracy of the weak-coupling approximation used in deriving Eq.~(\ref{Hk}). 
First define $\eta = (\bm{a},\bm{b})$ as being a list containing all fermionic operators of chains 1 and 2. 
Then define the $2L\times 2L$ covariance matrix
\begin{equation}\label{theta}
\theta_{i,j} = \langle \eta_i^\dagger \eta_j\rangle
\end{equation}
From the von Neumann equation it can be shown that
\begin{equation}\label{theta_time}
\frac{\ud \theta}{\ud t} = i [R,\theta]
\end{equation}
where
\begin{equation}\label{R}
R = \begin{pmatrix}
\Lambda & G \\
G\trans & \Lambda 
\end{pmatrix}
\end{equation}
with $\Lambda = \text{diag}(\lambda_k)$ and $G$ being the matrix with entries $G_{k,q}$ given in Eq.~(\ref{Gkl}). 
The initial conditions of Eq.~(\ref{theta_time}) correspond to the two chains being separately prepared in thermal equilibrium at temperatures $T_1$ and $T_2$. 
That is, $\theta(0) = \text{diag}(n_k^1, n_k^2)$ [cf. Eq.~(\ref{nk})]. 
The formal solution of Eq.~(\ref{theta_time}) is then
\begin{equation}\label{theta_sol}
\theta(t) = e^{i R t} \theta(0) e^{-i R t}
\end{equation}

We may use Eq.~(\ref{theta_sol}) to compute the average heat or any other expectation value and compare it with our approximate solution. 
A direct comparison with the average heat in Eq.~(\ref{Qave}) is given in Fig.~\ref{fig:NumExact1} for several sizes. 
It can be seen that the agreement in the ``well-behaved'' region is already very good for $L = 100$, and clearly improves with increasing size. 
In the unstable region the agreement is not as good, but the general features are unchanged. 
This agreement is not restrictive only to the average heat. The time dependence of all individual modes are also well approximated. 
In Fig.~\ref{fig:NumExact2} we compare the numerically exact solution for $\langle a_k^\dagger a_k \rangle$ with Eq.~(\ref{aa}) and find once again that, as the size increases, the general time dependence indeed converge to our approximate results.

%
%
%
%

\section{\label{sec:work}The statistics of work}

In Ref.~\cite{Apollaro2014} the authors studied the statistics of the work that must be performed in suddenly  turning on the interaction between two XX chains.
However, they used an exact diagonalization procedure and therefore only studied chains of small sizes.
Since the setup is identical to the one here, we may apply the same formalism to this problem. 
For simplicity I will henceforth assume both temperatures are equal. 

As with heat, computing the probability of work requires two measurements, one of $H_0$ and the other of $H = H_0 + V$ \cite{Campisi2011}. 
Let  $H|f\rangle = \mathcal{E}_f |f\rangle$. 
The corresponding distribution of work is then \cite{Talkner2007}
\begin{equation}\label{PW}
P(W) = \sum\limits_{\bm{n},f}  |\langle f  | \bm{n}\rangle|^2 \;p_{\bm{n}}\;\; \delta[W - (\mathcal{E}_f - (E_{n_1}+E_{n_2}))]
\end{equation}
and the characteristic function is \cite{Campisi2011}
\begin{equation}\label{MW}
M(u) = \langle e^{i u W} \rangle = \tr \left\{ e^{i u H} e^{-i u(H_1+H_2)} \rho_\text{th}\right\}
\end{equation}
This function obeys the Jarzynski equality \cite{Jarzynski1997a}.

Using the results of Sec.~\ref{sec:model} we find that $M(u)$ may also be written as a product:
\begin{equation}\label{Mk}
M(u) = \prod\limits_k M(k,u) = \prod\limits_k \Bigg\{ 1+ 2n_k(1-n_k) (\cos(G_k u)-1)\Bigg\}
\end{equation}
So, as before, we may speak of the work $W_k$ performed in turning on the coupling between the modes $a_k$ and $b_k$. 
This work may take on the values $\pm G_k$ and 0. Moreover the probabilities of positive and negative work are equal and have the value $n_k (1-n_k)$. 
Consequently, the average work performed is zero, a fact which can also be inferred simply from the symmetry of the interaction $V$. 

The total work is a sum of independent contributions and may therefore have any value consisting of combinations of the $G_k$. 
The maximum work possible is thus
\[
\sum\limits_k |G_k| = g_0
\]
which is  independent of $L$. 

\begin{figure}
\centering
\includegraphics[width=0.45\textwidth]{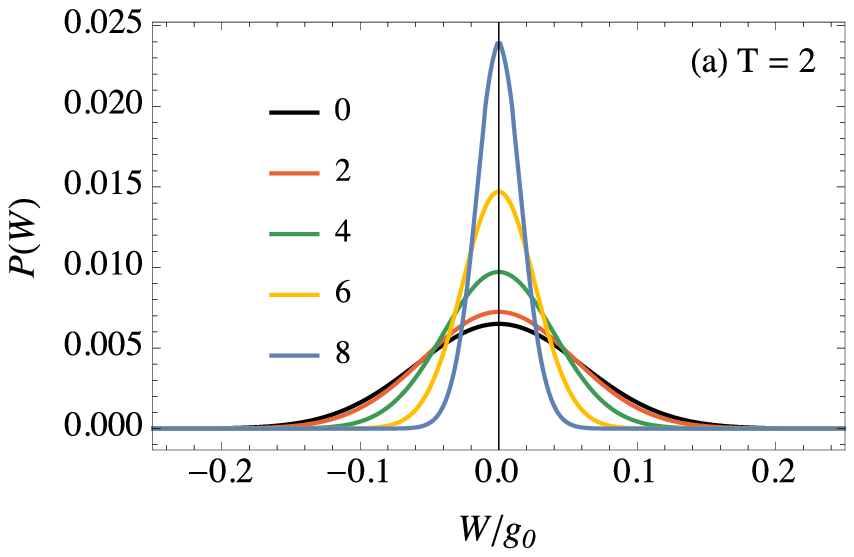}\\
\includegraphics[width=0.45\textwidth]{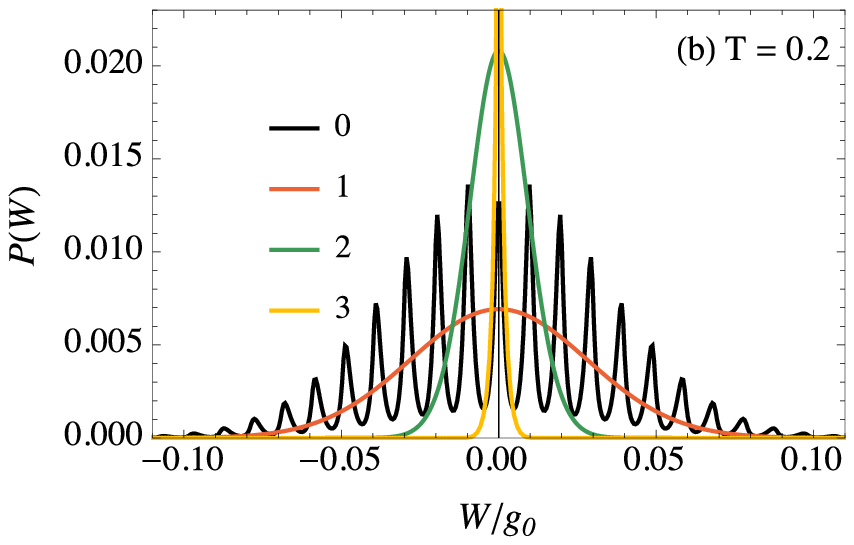}
\caption{\label{fig:WD} 
Work distribution for (a) $T = 2$ and (b) $T = 0.2$, with $L = 200$, $g_0= 0.1$ and several values of $h$, as indicated in each image.}
\end{figure}

\begin{figure}
\centering
\includegraphics[width=0.45\textwidth]{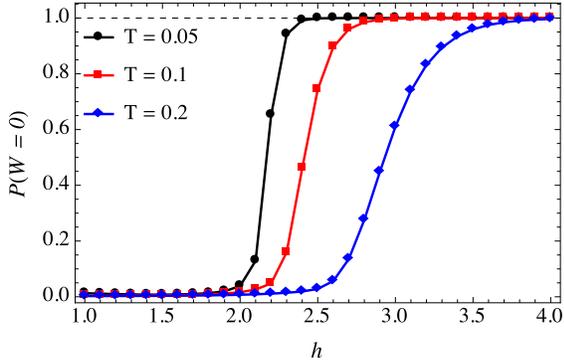}
\caption{\label{fig:W0} 
The probability that no work is performed as a function of $h$ for different temperatures with $L = 400$ and $g_0 = 0.1$. 
}
\end{figure}

Results for the work distribution, computed using the FFT method for $L = 200$, are shown in Fig.~\ref{fig:WD}. 
As can be seen,  the distribution for high temperatures is already at a Gaussian shape. 
However, at low temperatures the strong effect of the quantum phase transition causes the distribution to display a series of sharp peaks, which are strongly suppressed when the field increases.
In Fig.~\ref{fig:W0} we also present the probability that no work is performed. 
It can be seen quite clearly that at the critical field  $h_c = 2$ there is a sharp transition toward a state where $P(W=0) \simeq 1$.

\bibliography{/Users/gtlandi/Documents/library}
\end{document}